\documentclass[twoside,leqno]{article}
\usepackage{amsmath}
\usepackage{amsthm}
\usepackage{amssymb}
\usepackage{amscd}
\usepackage{graphicx}
\usepackage{epsfig}

\usepackage{latexsym}
\usepackage{amsfonts}

\input xy
\xyoption{all} \tolerance=500

-\headheight\advance\topmargin by -\headsep

\numberwithin{equation}{section} \allowdisplaybreaks

\newtheorem{theo}{Theorem}

\newtheorem{ex}{Example}
\newtheorem{Definition}{Definition}

\begin{document}
\font\black=cmbx10 \font\sblack=cmbx7 \font\ssblack=cmbx5
\font\blackital=cmmib10  \skewchar\blackital='177
\font\sblackital=cmmib7 \skewchar\sblackital='177
\font\ssblackital=cmmib5 \skewchar\ssblackital='177
\font\sanss=cmss10 \font\ssanss=cmss8 
\font\sssanss=cmss8 scaled 600 \font\blackboard=msbm10
\font\sblackboard=msbm7 \font\ssblackboard=msbm5
\font\caligr=eusm10 \font\scaligr=eusm7 \font\sscaligr=eusm5
\font\blackcal=eusb10 \font\fraktur=eufm10 \font\sfraktur=eufm7
\font\ssfraktur=eufm5 \font\blackfrak=eufb10

\font\bsymb=cmsy10 scaled\magstep2
\def\all#1{\setbox0=\hbox{\lower1.5pt\hbox{\bsymb
       \char"38}}\setbox1=\hbox{$_{#1}$} \box0\lower2pt\box1\;}
\def\exi#1{\setbox0=\hbox{\lower1.5pt\hbox{\bsymb \char"39}}
       \setbox1=\hbox{$_{#1}$} \box0\lower2pt\box1\;}

\def\mi#1{{\fam1\relax#1}}
\def\tx#1{{\fam0\relax#1}}

\newfam\bifam
\textfont\bifam=\blackital \scriptfont\bifam=\sblackital
\scriptscriptfont\bifam=\ssblackital
\def\bi#1{{\fam\bifam\relax#1}}

\newfam\blfam
\textfont\blfam=\black \scriptfont\blfam=\sblack
\scriptscriptfont\blfam=\ssblack
\def\rbl#1{{\fam\blfam\relax#1}}

\newfam\bbfam
\textfont\bbfam=\blackboard \scriptfont\bbfam=\sblackboard
\scriptscriptfont\bbfam=\ssblackboard
\def\bb#1{{\fam\bbfam\relax#1}}

\newfam\ssfam
\textfont\ssfam=\sanss \scriptfont\ssfam=\ssanss
\scriptscriptfont\ssfam=\sssanss
\def\sss#1{{\fam\ssfam\relax#1}}

\newfam\clfam
\textfont\clfam=\caligr \scriptfont\clfam=\scaligr
\scriptscriptfont\clfam=\sscaligr
\def\cl#1{{\fam\clfam\relax#1}}

\newfam\frfam
\textfont\frfam=\fraktur \scriptfont\frfam=\sfraktur
\scriptscriptfont\frfam=\ssfraktur
\def\fr#1{{\fam\frfam\relax#1}}

\def\cb#1{\hbox{$\fam\gpfam\relax#1\textfont\gpfam=\blackcal$}}

\def\hpb#1{\setbox0=\hbox{${#1}$}
    \copy0 \kern-\wd0 \kern.2pt \box0}
\def\vpb#1{\setbox0=\hbox{${#1}$}
    \copy0 \kern-\wd0 \raise.08pt \box0}

\def\pmb#1{\setbox0\hbox{${#1}$} \copy0 \kern-\wd0 \kern.2pt \box0}
\def\pmbb#1{\setbox0\hbox{${#1}$} \copy0 \kern-\wd0
      \kern.2pt \copy0 \kern-\wd0 \kern.2pt \box0}
\def\pmbbb#1{\setbox0\hbox{${#1}$} \copy0 \kern-\wd0
      \kern.2pt \copy0 \kern-\wd0 \kern.2pt
    \copy0 \kern-\wd0 \kern.2pt \box0}
\def\pmxb#1{\setbox0\hbox{${#1}$} \copy0 \kern-\wd0
      \kern.2pt \copy0 \kern-\wd0 \kern.2pt
      \copy0 \kern-\wd0 \kern.2pt \copy0 \kern-\wd0 \kern.2pt \box0}
\def\pmxbb#1{\setbox0\hbox{${#1}$} \copy0 \kern-\wd0 \kern.2pt
      \copy0 \kern-\wd0 \kern.2pt
      \copy0 \kern-\wd0 \kern.2pt \copy0 \kern-\wd0 \kern.2pt
      \copy0 \kern-\wd0 \kern.2pt \box0}

\def\cdotss{\mathinner{\cdotp\cdotp\cdotp\cdotp\cdotp\cdotp\cdotp
        \cdotp\cdotp\cdotp\cdotp\cdotp\cdotp\cdotp\cdotp\cdotp\cdotp
        \cdotp\cdotp\cdotp\cdotp\cdotp\cdotp\cdotp\cdotp\cdotp\cdotp
        \cdotp\cdotp\cdotp\cdotp\cdotp\cdotp\cdotp\cdotp\cdotp\cdotp}}

\font\frak=eufm10 scaled\magstep1 \font\fak=eufm10 scaled\magstep2
\font\fk=eufm10 scaled\magstep3 \font\scriptfrak=eufm10
\font\tenfrak=eufm10


\mathchardef\za="710B  
\mathchardef\zb="710C  
\mathchardef\zg="710D  
\mathchardef\zd="710E  
\mathchardef\zve="710F 
\mathchardef\zz="7110  
\mathchardef\zh="7111  
\mathchardef\zvy="7112 
\mathchardef\zi="7113  
\mathchardef\zk="7114  
\mathchardef\zl="7115  
\mathchardef\zm="7116  
\mathchardef\zn="7117  
\mathchardef\zx="7118  
\mathchardef\zp="7119  
\mathchardef\zr="711A  
\mathchardef\zs="711B  
\mathchardef\zt="711C  
\mathchardef\zu="711D  
\mathchardef\zvf="711E 
\mathchardef\zq="711F  
\mathchardef\zc="7120  
\mathchardef\zw="7121  
\mathchardef\ze="7122  
\mathchardef\zy="7123  
\mathchardef\zf="7124  
\mathchardef\zvr="7125 
\mathchardef\zvs="7126 
\mathchardef\zf="7127  
\mathchardef\zG="7000  
\mathchardef\zD="7001  
\mathchardef\zY="7002  
\mathchardef\zL="7003  
\mathchardef\zX="7004  
\mathchardef\zP="7005  
\mathchardef\zS="7006  
\mathchardef\zU="7007  
\mathchardef\zF="7008  
\mathchardef\zW="700A  

\newcommand{\be}{\begin{equation}}
\newcommand{\ee}{\end{equation}}
\newcommand{\ra}{\rightarrow}
\newcommand{\lra}{\longrightarrow}
\newcommand{\bea}{\begin{eqnarray}}
\newcommand{\eea}{\end{eqnarray}}
\newcommand{\beas}{\begin{eqnarray*}}
\newcommand{\eeas}{\end{eqnarray*}}
\def\*{{\textstyle *}}
\newcommand{\R}{{\mathbb R}}
\newcommand{\T}{{\mathbb T}}
\newcommand{\C}{{\mathbb C}}
\newcommand{\unit}{{\mathbf 1}}
\newcommand{\SL}{SL(2,\C)}
\newcommand{\Sl}{sl(2,\C)}
\newcommand{\SU}{SU(2)}
\newcommand{\su}{su(2)}
\def\ssT{\sss T}
\newcommand{\G}{{\goth g}}
\newcommand{\D}{{\rm d}}
\newcommand{\Df}{{\rm d}^\zF}
\newcommand{\de}{\,{\stackrel{\rm def}{=}}\,}
\newcommand{\we}{\wedge}
\newcommand{\nn}{\nonumber}
\newcommand{\ot}{\otimes}
\newcommand{\s}{{\textstyle *}}
\newcommand{\ts}{T^\s}
\newcommand{\oX}{\stackrel{o}{X}}
\newcommand{\oD}{\stackrel{o}{D}}
\newcommand{\obD}{\stackrel{o}{\bD}}
\newcommand{\pa}{\partial}
\newcommand{\ti}{\times}
\newcommand{\A}{{\cal A}}
\newcommand{\Li}{{\cal L}}
\newcommand{\ka}{\mathbb{K}}
\newcommand{\find}{\mid}
\newcommand{\ad}{{\rm ad}}
\newcommand{\rS}{]^{SN}}
\newcommand{\rb}{\}_P}
\newcommand{\p}{{\sf P}}
\newcommand{\h}{{\sf H}}
\newcommand{\X}{{\cal X}}
\newcommand{\I}{\,{\rm i}\,}
\newcommand{\rB}{]_P}
\newcommand{\Ll}{{\pounds}}
\def\lna{\lbrack\! \lbrack}
\def\rna{\rbrack\! \rbrack}
\def\rnaf{\rbrack\! \rbrack_\zF}
\def\rnah{\rbrack\! \rbrack\,\hat{}}
\def\lbo{{\lbrack\!\!\lbrack}}
\def\rbo{{\rbrack\!\!\rbrack}}
\def\lan{\langle}
\def\ran{\rangle}
\def\zT{{\cal T}}
\def\tU{\tilde U}
\def\ati{{\stackrel{a}{\times}}}
\def\sti{{\stackrel{sv}{\times}}}
\def\aot{{\stackrel{a}{\ot}}}
\def\sati{{\stackrel{sa}{\times}}}
\def\saop{{\stackrel{sa}{\op}}}
\def\bwa{{\stackrel{a}{\bigwedge}}}
\def\svop{{\stackrel{sv}{\oplus}}}
\def\saot{{\stackrel{sa}{\otimes}}}
\def\cti{{\stackrel{cv}{\times}}}
\def\cop{{\stackrel{cv}{\oplus}}}
\def\dra{{\stackrel{\xd}{\ra}}}
\def\bdra{{\stackrel{\bd}{\ra}}}
\def\bAff{\mathbf{Aff}}
\def\Aff{\sss{Aff}}
\def\bHom{\mathbf{Hom}}
\def\Hom{\sss{Hom}}
\def\bt{{\boxtimes}}
\def\sot{{\stackrel{sa}{\ot}}}
\def\bp{{\boxplus}}
\def\op{\oplus}
\def\bwak{{\stackrel{a}{\bigwedge}\!{}^k}}
\def\aop{{\stackrel{a}{\oplus}}}
\def\ix{\operatorname{i}}
\def\V{{\cal V}}
\def\cD{{\cal D}}
\def\cC{{\cal C}}
\def\cE{{\cal E}}
\def\cL{{\cal L}}
\def\cN{{\cal N}}
\def\cR{{\cal R}}
\def\cJ{{\cal J}}
\def\cT{{\cal T}}
\def\cH{{\cal H}}
\def\bA{\mathbf{A}}
\def\bI{\mathbf{I}}
\def\wh{\widehat}
\def\wt{\widetilde}
\def\ol{\overline}
\def\ul{\underline}
\def\Sec{\sss{Sec}}
\def\Lin{\sss{Lin}}
\def\ader{\sss{ADer}}
\def\ado{\sss{ADO^1}}
\def\adoo{\sss{ADO^0}}
\def\AS{\sss{AS}}
\def\bAS{\sss{AS}}
\def\bLS{\sss{LS}}
\def\bAP{\sss{AV}}
\def\bLP{\sss{LP}}
\def\AP{\sss{AP}}
\def\LP{\sss{LP}}
\def\LS{\sss{LS}}
\def\Z{\mathbf{Z}}
\def\oZ{\overline{\bZ}}
\def\oA{\overline{\bA}}
\def\cim{{C^\infty(M)}}
\def\de{{\cal D}^1}
\def\la{\langle}
\def\ran{\rangle}
\def\by{{\bi y}}
\def\bs{{\bi s}}
\def\bc{{\bi c}}
\def\bd{{\bi d}}
\def\bh{{\bi h}}
\def\bD{{\bi D}}
\def\bY{{\bi Y}}
\def\bX{{\bi X}}
\def\bL{{\bi L}}
\def\bV{{\bi V}}
\def\bW{{\bi W}}
\def\bS{{\bi S}}
\def\bT{{\bi T}}
\def\bC{{\bi C}}
\def\bE{{\bi E}}
\def\bF{{\bi F}}
\def\bP{{\bi P}}
\def\bp{{\bi p}}
\def\bz{{\bi z}}
\def\bZ{{\bi Z}}
\def\bq{{\bi q}}
\def\bQ{{\bi Q}}
\def\bx{{\bi x}}

\def\sA{{\sss A}}
\def\sC{{\sss C}}
\def\sD{{\sss D}}
\def\sG{{\sss G}}
\def\sH{{\sss H}}
\def\sI{{\sss I}}
\def\sJ{{\sss J}}
\def\sK{{\sss K}}
\def\sL{{\sss L}}
\def\sO{{\sss O}}
\def\sP{{\sss P}}
\def\sPh{{\sss P\sss h}}
\def\sT{{\sss T}}
\def\sV{{\sss V}}
\def\sR{{\sss R}}
\def\sS{{\sss S}}
\def\sE{{\sss E}}
\def\sF{{\sss F}}
\def\st{{\sss t}}
\def\sg{{\sss g}}
\def\sx{{\sss x}}
\def\sv{{\sss v}}
\def\sw{{\sss w}}
\def\sQ{{\sss Q}}
\def\sj{{\sss j}}
\def\sq{{\sss q}}
\def\xa{\tx{a}}
\def\xc{\tx{c}}
\def\xd{\tx{d}}
\def\xi{\tx{i}}
\def\xD{\tx{D}}
\def\xV{\tx{V}}
\def\xF{\tx{F}}
\def\dt{\xd_{\sss T}}
\def\dte{\dt^\ze}
\def\dtwte{\dt^{\wt{\ze}}}
\def\vt{\textsf{v}_{\sss T}}
\def\vta{\operatorname{v}_\zt}
\def\vtb{\operatorname{v}_\zp}
\def\cM{{\cal M}}
\def\cN{{\cal N}}
\def\cD{{\cal D}}
\def\ug{{\ul{\zg}}}
\def\rel{{-\!\!\!-\!\!\rhd}}
\newdir{|>}{%
!/4.5pt/@{|}*:(1,-.2)@^{>}*:(1,+.2)@_{>}}

\setcounter{page}{1} \thispagestyle{empty}


\bigskip

\bigskip

\title{Variational calculus with constraints on general algebroids}

        \author{
        Katarzyna  Grabowska$^1$, Janusz Grabowski$^2$\thanks{Research financed by the Polish
Ministry of Science and Higher Education under the grant No. N201 005 31/0115.}\\
        \\
         $^1$ {\it Physics Department}\\
                {\it University of Warsaw} \\
         $^2$ {\it Institute of Mathematics}\\
                {\it Polish Academy of Sciences}
                }
\date{}
\maketitle
\begin{abstract}
Variational calculus on a vector bundle $E$ equipped with a structure of a general
algebroid is developed, together with the corresponding analogs of Euler-Lagrange
equations. Constrained systems are introduced in the variational and in the geometrical
setting. The constrained Euler-Lagrange equations are derived for analogs of holonomic,
vakonomic and nonholonomic constraints. This general model covers majority of
first-order Lagrangian systems which are present in the literature and reduces to the
standard variational calculus and the Euler-Lagrange equations in Classical Mechanics
for $E=TM$.


\end{abstract}
\section{Introduction}
The Classical Analytical Mechanics is an old and well-established
part of both mathematics and physics. Nevertheless many people
still look for the best mathematical tools in describing various
aspects of mechanical systems. A use of Lie algebroids and Lie
groupoids for describing some systems of the Classical Mechanics
was proposed by P.~Libermann \cite{Li} and A.~Weinstein \cite{We}
more than ten years ago. This turned out to be a very fruitful
idea and since then much work has been done (e.g.
\cite{CLMMM,GGU3,LMM,IMMS,IMPS,Mar1,M1}) making use of Lie
algebroids in various aspects of Classical Mechanics and Classical
Field Theory. The need of extending the geometrical tools of the
Lagrangian formalism from just tangent bundles to Lie algebroids
is justified by the fact that reductions usually move us out of
the environment of the tangent bundles (think on the rigid body).
It is similar to the better-known situation of passing from the
symplectic to the Poisson structures in the Hamiltonian formalism.

In the paper \cite{GGU3} it was observed that, following some ideas of W.~M.~Tulczyjew and using general
algebroids instead of just Lie algebroids, one can describe a larger class of systems in a simple and elegant
way, both in the Lagrangian and in the Hamiltonian formulation. Moreover, the proposed geometric picture does
not require considering prolongations of Lie algebroids we start with, as it was in the case of previous
approaches known in the literature. A further paper \cite{GGU4}, was devoted, in turn, to the construction of
Euler-Lagrange equations in the affine setting of so called {\it special affgebroids} which is particularly
suitable for time-dependent systems.

In the present paper we concentrate on variational calculus and constraints in the algebroid setting. We work
with a general algebroid, defined in \cite{GU2} as a double vector bundle morphism
\be\label{tt}\varepsilon:\sT^\ast E\rightarrow \sT E^\ast\ee
covering the identity on $E^\ast$. Here $\tau: E\rightarrow M$ is a vector bundle
playing the role of kinematic configurations. To some extent then, our paper can be
understood as a natural generalization of \cite{Mar3}, where a variational calculus on
Lie algebroids has been developed according to the original ideas of A.~Weinstein
\cite{We}, and of \cite{CLMM,IMMS}, where constraints on Lie algebroids have been
considered. On the other hand, our approach is definitely different from the approaches
known in the literature, even when the equations we obtain cover the corresponding
Euler-Lagrange equations in the Lie algebroid case. This is mainly because we adapt the
framework of the {\it Tulczyjew triple} \cite{Tu1,Tu3,TU}, working simply with the
morphism (\ref{tt}) rather than following the Klein's method \cite{Kl} generalized to
Lie algebroids, in which the bundles tangent to $E$ and $E^\*$ are replaced by the
prolongations of $E$ with respect to the vector bundle projections $\zt\colon E
\rightarrow M$ and $\zp\colon E^\* \rightarrow M$. This, in our opinion, simplifies the
whole formalism substantially.

To define a variational problem on an algebroid we have to specify a manifold $\mathcal M$ of paths whose
tangent space $\sT\mathcal M$ represents all possible variations and an action functional $W$ on $\mathcal M$.
Then we have to choose a submanifold $\cN$ of {\it admissible paths} and a set (generalized distribution)
$\cD\subset\sT\cM_{\mid\cN}$ of {\it admissible variations} of admissible paths. In \cite{Mar3} admissible
variations are constructed out of homotopies of admissible paths as defined in \cite{CF}. For general
algebroids we need different way of constructing admissible variations, since we have to accept the fact that
they are not tangent to the submanifold of admissible paths in general. Therefore we construct admissible
variations for an admissible path $\zg$ in $E$ out of {\it vertical variations} of $\zg$ in $E$, i.e. out of
vertical vector fields along $\zg$. Note that the variations are defined in $E$ (which is $\sT M$ in the
standard variational calculus), not in $M$. This is because the variational calculus on algebroids leads to
first-order differential equations in $E$ rather than to second-order equations in $M$. This is only the case
of the canonical Lie algebroid $E=\sT M$ when paths in $M$ are in one to one correspondence with admissible
paths in $E$, this time -- just tangent prolongations of paths in $M$, and admissible variations are tangent
prolongations of variations of paths in $M$. For a general algebroid the admissible variations are constructed
from the vertical ones by means of the double vector bundle relation $\zk=\zk_\ze:\sT E\rel\sT E$ which is
dual to the morphism $\varepsilon$. Of course, for Lie algebroids our admissible paths coincide with the
infinitesimal homotopies of admissible paths associated with the lifts of time-dependent sections, as they
appear in \cite{Mar3,CF}. We prefer a more fundamental approach which uses $\zk_\ze$ to produce admissible
variations out of the vertical ones, instead of lifting whole sections extending paths in $E$ and showing that
the result does not depend on the extension. In the case of $E=\sT M$ the mapping $\varepsilon$ defining an
algebroid structure is the inverse to the Tulczyjew isomorphism $\alpha_M:\sT\sT^\ast M\ra\sT^\ast\sT M$. The
relation $\kappa_\varepsilon$ is in this case the well-known canonical flip $\kappa_M: \sT\sT M\rightarrow
\sT\sT M$. Our construction is especially convenient in the case of non-holonomic constraints where
variations are not tangent to the submanifold of constraints.

It is clear from our variational picture that putting constraints must result in defining a subset of
$\mathcal D$. In the case of a general algebroid $E$ our classification of the constraints is based on the way
in which the constrained admissible variations are constructed. According to the tradition we call them:
vakonomic, non-holonomic, and holonomic constraints. Starting from a subset $S$ of $E$, classically understood
as a geometric constraint for velocities, we have at least two natural possibilities of constructing a
constraint in admissible variations: one is to consider only admissible variations which are tangent to $S$
({\it vakonomic constraint}), the other -- to consider only admissible variations coming from those vertical
ones which are tangent to $S$ ({\it nonholonomic constraint}). Note that our approach allows to understand
nonholonomic constraint as a constrained variational problem, contrary to the commonly accepted conviction. A
nonholonomic constraint is called {\it holonomic} if the constrained admissible variations are tangent to $S$
(are vakonomic). Sometimes it is hard to decide without making an experiment which method should be used to
describe the real behavior of the system.

For all types of constraints we construct analogs of the Euler-Lagrange equation for systems that are subject
to those three types of constraints in variational way. Note however that the corresponding equations describe
"regular" solutions rather than a general solution of the variational problem. Additionally, like for
non-constrained cases in \cite{GGU3}, we derive the equations purely geometrically, without referring to the
variational calculus.

The literature concerning constraints in Variational Calculus is so extensive that there it is impossible to
cite it in a complete way. We decided to list among references only papers dealing actually with Lie
algebroids or being direct inspiration for the framework we propose. Let us also make it clear that we see the
meaning of the present paper not only as a generalization of formalisms of Classical Mechanics. Working with
the case of a general algebroid forced us to propose a geometric approach which seems to be new and
illustrative even when applied to very classical situations. The main observation is that an algebroid
structure is a crucial geometric ingredient in constructing the dynamics of the system. It tells us not only
the configurations, velocities and inner degrees of freedom, but it contains the information on how the
admissible variations should be produced from a simple geometric model of variations of paths in a vector
bundle -- the vertical ones. This structure is encoded in a single map (\ref{tt}) respecting double vector
bundle structures. The brackets and the Jacobi identity are therefore proven to play a minor role. The Jacobi
identity for an algebroid bracket ensures some integrability conditions that allow us to integrate the Lie
algebroid into an (at least local) Lie groupoid (see \cite{CF}), but which is irrelevant for the possibility
of constructing Euler-Lagrange equations. Fixing this geometric setting for our system, it is then the
Lagrangian function which produces a concrete dynamics out of these data. However, we would like to stress
that regularity of the Lagrangian is completely irrelevant for our picture. The general method of constructing
dynamics out of the Lagrangian works for all Lagrangians, singular or not. The difficulty with singular
Lagrangians is that the dynamics we obtain is really implicit and complicated. In other words, difficulty with
singular Lagrangians lies in difficulty in solving equations, not in the geometric construction of the
equations themselves.

Finally, if the variational calculus is concerned, only admissible paths come to the play. This is because we
work on the bundle $E$ of kinematic configurations and considering only admissible paths corresponds,
classically, to work with paths in the manifold $M$ of position configurations lifted canonically to the paths
in $\sT M$. The geometrical model of (infinitesimal) variations of an admissible path $\zg$ is to consider
vertical vector fields along $\zg:[t_0,t_1]\ra E$. Now, the true (mechanical) admissible variations are vector
fields along $\zg$ constructed from the vertical ones out of the algebroid structure $\zk$. This is how the
algebroid structure comes to the variational picture. Note that the role of the (Lie) algebroid structure in
the classical setting is usually overlooked, since it is hidden behind structures of the tangent and cotangent
bundles which are viewed as a natural part of the theory.

The paper is organized as follows. In Section 2 we set up the notation and we recall the notion of general
algebroid as a double vector bundle morphism. Then we introduce the relation $\kappa$ that is used for
defining admissible variations. In Section 3 we discuss the Lagrange formalism without constraints on general
algebroid. Then we pass in Section 4 to the variational calculus. We derive the variation of the Lagrangian
and Euler-Lagrange equations. The final section is devoted to constraints. Geometric constraints as subsets
$S\subset E$ give rise to variational constraints which are classified in pure geometrical terms as vakonomic,
nonholonomic, or holonomic. We derive constrained equations using variational motivations and give them pure
geometric interpretations.

\section{Lie algebroids as double vector bundle morphisms\label{S1}}
We start with introducing some notation.

Let $M$ be a smooth manifold and let $(x^a), \ a=1,\dots,n$, be a coordinate system in $M$. We denote by
$\zt_M \colon \sT M \rightarrow M$ the tangent vector bundle and by $\zp_M \colon \sT^\* M\rightarrow M$ the
cotangent vector bundle. We have the induced (adapted) coordinate systems $(x^a, {\dot x}^b)$ in $\sT M$ and
$(x^a, p_b)$ in $\sT^\* M$.
        Let $\zt\colon E \rightarrow M$ be a vector bundle and let $\zp
\colon E^\* \rightarrow M$ be the dual bundle.
  Let $(e_1,\dots,e_m)$  be a basis of local sections of $\zt\colon
E\rightarrow M$ and let $(e^{1}_*,\dots, e^{m}_*)$ be the dual basis of local sections of $\zp\colon
E^\*\rightarrow M$. We have the induced coordinate systems:
    \beas
    (x^a, y^i),\quad & y^i=\zi(e^{i}_*), \quad \text{in} \ E,\\
    (x^a, \zx_i), \quad &\zx_i = \zi(e_i),\quad \text{in} \ E^\* ,
    \eeas
    where the linear functions  $\zi(e)$ are given by the canonical pairing
    $\zi(e)(v_x)=\la e(x),v_x\ran$. Thus we have local coordinates
    \beas
    (x^a, y^i,{\dot x}^b, {\dot y}^j ) &  \quad \text{in} \ \sT E,\\
    (x^a, \zx_i, {\dot x}^b, {\dot \zx}_j) & \quad \text{in} \ \sT E^\* ,\\
    (x^a, y^i, p_b, \zp_j) & \quad \text{in}\ \sT^\*E,\\
    (x^a, \zx_i, p_b, \zf^j) & \quad \text{in}\ \sT^\* E^\* .
    \eeas

It is well known (cf. \cite{KU,Ur}) that the cotangent bundles $\sT^\*E$ and $\sT^\*E^\*$ are examples of
double vector bundles: $$\xymatrix{
\sT^\ast E^\ast\ar[rr]^{\sT^\ast\zp} \ar[d]_{\zp_{E^\ast}} && E\ar[d]^{\zt} \\
E^\ast\ar[rr]^{\zp} && M } \qquad {,}\qquad \xymatrix{
\sT^\ast E\ar[rr]^{\sT^\ast\zt} \ar[d]_{\zt_{E^\ast}} && E^\ast\ar[d]^{\zp} \\
E\ar[rr]^{\zt} && M }.
$$
Note that the concept of a double vector bundle goes back to J.~Pradines \cite{Pr1,Pr2}, see also
\cite{Ma,KU}. In particular, all arrows correspond to vector bundle structures and all pairs of vertical and
horizontal arrows are vector bundle morphisms. The double vector bundles have been recently characterized
\cite{GR} in a simple way as two vector bundle structures whose Euler vector fields commute. The above double
vector bundles are canonically isomorphic with the isomorphism
    \be\label{iso}\cR_\zt \colon \sT^\*E \longrightarrow \sT^\* E^\*
                    \ee
  being simultaneously an anti-symplectomorphism  (cf. \cite{Du,KU,GU2}). In local coordinates, $\cR_\zt$ is given by
    $$\cR_\zt(x^a, y^i, p_b, \zp_j) = (x^a, \zp_i, -p_b,y^j).
                              $$
This means that we can identify coordinates $\zp_j$ with $\zx_j$, coordinates $\zf^j$
with $y^j$, and use the coordinates $(x^a, y^i, p_b, \zx_j)$ in $\sT^\ast E$ and the
coordinates $(x^a, \zx_i, p_b,y^j)$ in $\sT^\ast E^\ast$, in full agreement with
(\ref{iso}).

For the standard concept and theory of {\it Lie algebroids} we refer to the survey
article \cite{Ma0} (see also \cite{Gr,Mac}). It is well known that Lie algebroid
structures on a vector bundle $E$ correspond to linear Poisson tensors on $E^\*$. A
2-contravariant tensor $\zP$ on $E^\*$ is called {\it linear} if the corresponding
mapping $\widetilde{\zP} \colon \sT^\* E^\* \rightarrow \sT E^\*$ induced by
contraction, $\wt{\zP}(\zn)=i_\zn\zP$, is a morphism of double vector bundles. One can
equivalently say that the corresponding bracket of functions is closed on (fiber-wise)
linear functions. The commutative diagram
$$\xymatrix{
\sT^\ast E^\ast\ar[r]^{\widetilde\Pi}  & \sT E^\ast \\
\sT^\ast E\ar[u]_{\cR_\tau}\ar[ur]^{\ze} & },
$$
describes a one-to-one correspondence between linear 2-contravariant tensors $\zP$ on $E^\*$ and morphisms
$\ze$ (covering the identity on $E^\*$) of the following double vector bundles (cf. \cite{KU, GU2}) :

\be\xymatrix{
 & \sT^\ast E \ar[rrr]^{\varepsilon} \ar[dr]^{\pi_E}
 \ar[ddl]_{\sT^\ast\tau}
 & & & \sT E^\ast\ar[dr]^{\sT\pi}\ar[ddl]_/-20pt/{\tau_{E^\ast}}
 & \\
 & & E\ar[rrr]^/-20pt/{\zr}\ar[ddl]_/-20pt/{\tau}
 & & & \sT M \ar[ddl]_{\tau_M}\\
 E^\ast\ar[rrr]^/-20pt/{id}\ar[dr]^{\pi}
 & & & E^\ast\ar[dr]^{\pi} & &  \\
 & M\ar[rrr]^{id}& & & M &
}\label{F1.3}
\ee
In local coordinates, every  such $\ze$ is of the form
\be\label{F1.4}
\ze(x^a,y^i,p_b,\zx_j) = (x^a, \zx_i, \zr^b_k(x)y^k, c^k_{ij}(x) y^i\zx_k + \zs^a_j(x) p_a)
\ee
(summation convention is used) and it corresponds to the linear tensor $$ \zP_\ze =c^k_{ij}(x)\zx_k
\partial _{\zx_i}\otimes \partial _{\zx_j} + \zr^b_i(x) \partial _{\zx_i}
\otimes \partial _{x^b} - \zs^a_j(x)\partial _{x^a} \otimes \partial _{\zx_j}.
$$
The morphisms (\ref{F1.3}) of double vector bundles covering the identity on $E^\*$ has been called an {\it
algebroid} in \cite{GU2}, while a {\it Lie algebroid} has turned out to be an algebroids for which the tensor
$\zP_\ze$ is a Poisson tensor. We can consider the {\it adjoint tensor} $\zP_\ze^+$, i.e. the 2-contravariant
tensor obtained from $\zP_\ze$ by transposition:
$$ \zP_\ze^{+} =c^k_{ji}(x)\zx_k \partial _{\zx_i}\otimes \partial _{\zx_j} +
\zr^b_i(x) \partial _{x^b}\ot\partial _{\zx_i}  - \zs^a_j(x)
\partial _{\zx_j}\ot\partial _{x^a}
$$
and the {\it opposite tensor} $-\zP_\ze$. It is clear that $\zP_\ze^{+}$ and $-\zP_\ze$ are linear. They
correspond therefore to new algebroid structures:  the {\it adjoint algebroid structure} $\ze^{+}$ and the
{\it opposite algebroid structure} $\bar{\ze}$. An algebroid we call a {quasi-Lie algebroid} if
${\ze}^{+}=\bar{\ze}$.

The relation to the canonical definition of Lie algebroid is given by the following
theorem (cf. { \cite{GU3, GU2}}).

\begin{theo}
An algebroid structure $(E,\ze)$ can be equivalently defined as a bilinear bracket $[\cdot ,\cdot]_\ze $ on
the space $\Sec(E)$ of sections of $\zt\colon E\rightarrow M$, together with vector bundle morphisms $\zr,\,
\zs \colon E\rightarrow \sT M$ ({\it left anchor} and {\it right anchor}), such that
$$ [fX,gY]_\ze = f\cdot\zr(X)(g)Y -g\cdot\zs(Y)(f) X
+fg [X,Y]_\ze
$$
         for $f,g \in \cC^\infty (M)$, $X,Y\in \Sec(E)$.
The bracket and anchors are related to the  bracket $\{\zf,\psi\}_{\zP_\ze}=\la\zP_\ze,\xd\zf\ot\xd\psi\ran$
in the algebra of functions on $E^\ast$ which is associated with the 2-contravariant tensor $\zP_\ze$ by the
formulae
\beas
        \zi([X,Y]_\ze)&= \{\zi(X), \zi(Y)\}_{\zP_\ze},  \\
        \zp^\*(\zr(X)(f))       &= \{\zi(X), \zp^\*f\}_{\zP_\ze}, \\
        \zp^\*(\zs(X)(f))       &= \{\zp^\* f, \zi(X)\}_{\zP_\ze}.
                                                   \eeas
        The algebroid $(E,\ze)$ is a quasi-Lie algebroid if and only if the tensor
$\zP_\ze$ is skew-symmetric, and it is a Lie algebroid if and only if the tensor
$\zP_\ze$ is a Poisson tensor.
\end{theo}
Since the dual bundles of $\zp_E:\sT^\ast E\ra E$ and $\sT\zp:\sT E^\ast\ra\sT M$ are,
respectively, $\zt_E:\sT E\ra E$ and $\sT\zt:\sT E\ra\sT M$, the dual to $\ze$ is a
relation $\zk=\zk_\ze:\sT E\rel\sT E$. It is a uniquely defined smooth submanifold
$\zk$ in $\sT E\ti\sT E$ consisting of pairs $(v,v')$ such that
$\zr(\zt_E(v'))=\sT\zt(v)$ and
$$\la v,\ze(v^\ast)\ran_{\sT\zt}=\la v',v^\ast\ran_{\zt_E}$$
for any $v^\ast\in\sT^\ast_{\zt_E(v')}E$, where $\la \cdot,\cdot\ran_{\sT\zt}$ is the
canonical pairing between $\sT E$ and $\sT E^\ast$, and $\la \cdot,\cdot\ran_{\zt_E}$
is the canonical pairing between $\sT E$ and $\sT^\ast E$. We will write $\zk:v\rel\,
v'$ instead of $(v,v')\in\zk$. This relation can be put into the following diagram of
"double vector bundle relations"
\be\xymatrix{
 & \sT E  \ar[dr]^{\zt_E}
 \ar[ddl]_{\sT\tau}
 & & & \sT E\ar @{-|>}[lll]_{\zk}\ar[dr]^{\sT\zt}\ar[ddl]_/-20pt/{\tau_{E}}
 & \\
 & & E\ar[rrr]^/-20pt/{\zr}\ar[ddl]_/-20pt/{\tau}
 & & & \sT M \ar[ddl]_{\tau_M}\\
 \sT M\ar[dr]^{\zt_M}
 & & & E\ar[dr]^{\zt}\ar[lll]_{\zs} & &  \\
 & M\ar[rrr]^{id}& & & M &
}\label{rel}
\ee
The relation $$\xymatrix{
\sT E \ar[d]^{\zt_{E}} && \sT E\ar[d]^{\sT\zt}\ar @{-|>}[ll]_{\zk} \\
E\ar[rr]^{\zr} && \sT M }
$$
is a {\it vector bundle morphism of the second kind}, i.e. it is represented by linear maps of the fiber $\sT
E$ over $v\in\sT M$ into the fibers $\sT_eE$ for all $e\in E$ such that $\zr(e)=v$. This is also the simplest
example of a morphism of Lie groupoids in the sense introduced and exploited by S.~Zakrzewski \cite{Zak}. To
such relations we will refer therefore as to {\it Zakrzewski morphisms}. The expression of the Zakrzewski
morphism (\ref{rel}), dual to $\ze$, in local coordinates reads
\be\label{rellocal}\zk:\left(x^a,\,{Y}^i,\,\zr^b_k(x)y^k,\,\dot{Y}^j\right)\rel\left(x^a,y^i,\,\zs^b_k(x)Y^k,\,
\dot{Y}^j+c^j_{kl}(x)y^kY^l\right)\,.
\ee
It is easy to see that the relation $\zk^{-1}_\ze$ coincides with $\zk_{\bar{\ze}^{+}}$. Thus $\zk=\zk^{-1}$
for quasi-Lie algebroids.

A canonical example of a mapping $\ze$ in the case of $E=\sT M$ is given by $\ze = \ze_M = \za^{-1}_M$ -- the
inverse to the Tulczyjew isomorphism $\za_M:\sT\sT^\*M\ra\sT^\*\sT M$ \cite{Tu1}. The dual Zakrzewski morphism
is in this case the well-known `canonical flip' $\zk_M:\sT\sT M\ra\sT\sT M$. Since $\za_M$ is an isomorphism,
$\zk_M$ is a true map, in fact -- an isomorphism of the corresponding two vector bundle structures as well.

\medskip
A $C^1$-curve $\zg:\R\ra E$ (or a $C^1$-path $\zg:[t_0,t_1]\ra E$) in an algebroid $E$ we call {\it
admissible}, if the tangent prolongation $\st(\ul{\zg})$ of its projection $\ul{\zg}=\zt\circ\zg$ coincides
with its anchor:
\be\label{adm} \st(\ul{\zg})=\zr(\zg(t)).\ee A curve (path) in the
canonical Lie algebroid $\sT M$ is admissible if and only if it is a tangent prolongation of its projection on
$M$. If we denote $\sT^{hol}E$ the subset of $\sT E$ consisting of {\it holonomic vectors},
\be\label{hol}\sT^{hol}E=\{ v\in\sT E: \sT\zt(v)=\zr(\zt_E(v))\}\,,\ee then admissible
curves (paths) in the algebroid $E$ can be characterized as those curves (paths) whose tangent prolongations
lay in $\sT^{hol}E$. The set of holonomic vectors $\sT^{hol}E$ can be equivalently characterized as the subset
in $\sT E$ which is mapped {\it via} $\sT\zr:\sT E\ra\sT\sT M$ to classical holonomic vectors $\sT^2 M=\{
u\in\sT\sT M:\zk_M(u)=u\}$, that justifies the name. In other words,
$$\sT^{hol}E=(\sT\zr)^{-1}(\sT^2 M)\,.$$
Note also that $\sT^{hol}E$ is canonically an affine bundle over $E$ modelled on the vertical bundle $\sV
E\subset\sT E$. In local coordinates, $\sT^{hol}E$ as submanifold in $\sT E$ is characterized by the equations
$\dot{x}^a=\zr^a_i(x)y^i$, so $(x^a,y^i,\dot{y}^j)$ can serve as local coordinates in $\sT^{hol}E$. It is easy
to see that, for quasi-Lie algebroids, $\zk(\sT^{hol}E)=\sT^{hol}E$.

Let now $\zg:[t_0,t_1]\ra E$ be a path and $\zz:[t_0,t_1]\ra\sV E\subset\sT E$ be a vertical vector field
along $\zg$, $\zt_E(\zz(t))=\zg(t)$. It is well known that $\sV E\simeq E\oplus_M E$, so vertical vectors at
$e\in E$ can be canonically identified with vectors of the fibre $E_{\zt(e)}$. Thus, the vertical vector field
$\zz$ can be identified with a path $\zz_E$ in $E$ covering $\ul{\zg}$. We can consider now the tangent
prolongation $\st(\zz_E)$ to get a vector field along $\zz_E$. The operation $\zz\mapsto \st(\zz_E)$
associates with any path $\zz$ in $\sV E$ a path $\st(\zz_E)$in $\sT E$. For $v\in\sT E$, in turn, the family
$\zk(v)$ defines a vector field over $\zr^{-1}(\sT\zt(v))$. More precisely, for every
$e\in\zr^{-1}(\sT\zt(v))$ there is a unique vector $\zk(v)_e\in\sT_eE$ such that $\zk(v)_e\in\zk(v)$. We get
the following.
\begin{theo}
If $\zg:[t_0,t_1]\ra E$ is an admissible path in $E$, then every vertical vector field
$\zz:[t_0,t_1]\ra\sV E$ along ${\zg}$ defines canonically a vector field
$\zd_\zz\zg:[t_0,t_1]\ra\sT E$ along $\zg$ by
\be\label{tlift}\zd_\zz\zg(t)=\zk(\st(\zz_E)(t))_{\zg(t)}\,.\ee In local coordinates,
with $\zg(t)=(x^a(t),y^i(t))$ and $\zz(t)=(x^a(t),y^i(t),0,f^i(t))$,
\be\label{tlift1}\zd_\zz\zg(t)=f^j(t)\zs^b_j(x(t))\pa_{x^b}(\zg(t))+\left(\frac{\xd f^k}{\xd t}(t)
+c^k_{ij}(x(t))y^i(t)f^j(t)\right)\pa_{y^k}(\zg(t))\,.\ee In other words, in local coordinates in $\sT E$,
\be\label{tlift1a}\zd_\zz\zg(t)=\left(x^a(t),y^i(t),\,f^j(t)\zs^b_j(x(t)),\,\frac{\xd
f^k}{\xd t}(t) +c^k_{ij}(x(t))y^i(t)f^j(t)\right)\,.\ee
\end{theo}
\noindent The vertical vector fields $\zz$ along $\zg$ we will call {\it vertical
variations} or {\it vertical virtual displacements} of $\zg$ and the vector fields
$\zd_\zz\zg$ along $\zg$ -- {\it admissible variations} or {\it admissible virtual
displacements}. Note that the space $\sV({\zg})$ vertical variations of $\zg$ is
canonically an (infinite-dimensional) vector space.

\medskip\noindent
{\bf Remark.} In \cite{CF,Mar3}, analogs of the admissible variations $\zd_\zz\zg$ have
been obtained (in Lie algebroid context, of course) from tangent lifts of
time-dependent sections of $E$. The tangent lifts of sections have natural
generalizations for general algebroids \cite{GGU3,GGU2}. We prefer, however, to define
the admissible variation $\zd_\zz\zg$ directly by means of the vertical variation $\zz$
and the relation $\zk$, as being more fundamental and conceptually closer to the
standard concepts of variations.

\section{Lagrangian formalism for general
algebroids} The double vector bundle morphism (\ref{F1.3}) can serve as geometric background for generalized
Lagrangian formalisms.

The Lagrangian $L:E\ra{\R}$ defines two smooth maps: the {\em Legendre mapping}: {
$\lambda_L:E\longrightarrow E^\ast$, $\lambda_L=\tau_{E^\ast}\circ\ze\circ\xd L$,}
which is covered by the {\em Tulczyjew differential} { $\Lambda_{L}: E\longrightarrow
\sT E^\ast$, $\Lambda_{L}=\ze\circ\xd L$}:
\be\label{diag}\xymatrix{
\sT^\ast E\ar[rr]^{\ze}  && \sT E^\ast\ar[d]^{\zt_{E^\ast}} \\
E\ar[rr]^{\zl_L}\ar[u]^{\xd L}\ar@{.>}[rru]^{\Lambda_{L}} && E^\ast }.
\ee The lagrangian function $L$ defines therefore {\em the
phase dynamics} { $\zG=\Lambda_{L}(E)\subset \sT E^\ast$} which can be understood as an implicit differential
equation on $E^\*$, solutions of which are `phase trajectories' of the system $\zb:\R\ra E^\ast$ and satisfy
$\st(\zb)(t)\in \zG$. An analog of the Euler-Lagrange equation for curves {\ $\gamma: \R\rightarrow E$} is
then
$$(E_L):\qquad \st(\zl_L\circ\gamma)=\Lambda_L\circ\gamma.$$
The equation $(E_L)$ simply means that $\zL_L\circ\zg$ is an admissible curve in $\sT E^\ast$, thus it is the
tangent prolongation of $\zl_L\circ\zg$. In local coordinates, $\zG$ has the parametrization by $(x^a,y^k)$
via $\zL_L$ in the form (cf. (\ref{F1.4}))
\be\zL_L(x^a,y^i)= \left(x^a,\frac{\partial L}{\partial y^i}(x,y),
\zr^b_k(x)y^k, c^k_{ij}(x) y^i\frac{\partial L}{\partial y^k}(x,y) + \zs^a_j(x)\frac{\partial L}{\partial
x^a}(x,y)\right) \label{F1.4a}\ee and the equation $(E_L)$, for $\zg(t)=(x^a(t),y^i(t))$, reads
\be (E_L):\qquad\frac{\xd x^a}{\xd t}=\zr^a_k(x)y^k, \quad
\frac{\xd}{\xd t}\left(\frac{\partial L}{\partial y^j}\right)= c^k_{ij}(x) y^i\frac{\partial L}{\partial y^k}
+ \zs^a_j(x)\frac{\partial L}{\partial x^a},\label{EL2}\ee in the full agreement with \cite{LMM, Mar1, Mar2,
We}, if only one takes into account that, for Lie algebroids, $\zs^a_j=\zr^a_j$. As one can see from
(\ref{EL2}), the solutions are automatically admissible curves in $E$, i.e. $\zr(\zg(t))=\st(\zt\circ\zg)(t)$.
As a curve in the canonical Lie algebroid $\sT M$ is admissible if and only if it is a tangent prolongation of
its projection on $M$, first-order differential equations for admissible curves (paths) in $\sT M$ may be
viewed as certain second-order differential equations for curves (paths) in $M$. This explains why,
classically, the Euler-Lagrange equations are regarded as second-order equations.

\medskip\noindent
{\bf Remark.} The Tulczyjew differential $\zL_L:\sT M\ra\sT\sT^\ast M$ with a given
Lagrangian function $L$ on the canonical Lie algebroid $E=\sT M$ is sometimes called
the {\it time evolution operator $K$} (see \cite{BGPR}), as the first ideas of this
operator go back to a work by S.~Kamimura \cite{Ka}. This operator has been studied by
several authors in many variational contexts, however, without recognition of its
direct relation to a (Lie) algebroid structure. We named this map after
W.~M.~Tulczyjew, since our understanding is based on his ideas \cite{Tu3}.

\medskip
The time-dependent version of the above picture is the following. Consider the direct product $\wt{E}=E\ti\sT
\R$ of the algebroid $E$ with the canonical (Lie) algebroid $\sT\R$ equipped with canonical coordinates
$(t,\dot{t})$. The corresponding algebroid morphism is clearly the product of $\ze$ and the inverse of the
Tulczyjew isomorphism $\za_R$:
\be\label{t-dep}\wt{\ze}=(\ze,\za_\R^{-1}):\sT^\ast\wt{E}=\sT^\ast E\ti\sT^\ast\sT\R\ra
\sT E^\ast\ti\sT\sT^\ast\R=\sT\wt{E}^\ast\,.
\ee
The affine hyperbundle $\A_\R=\{(t,1)\in\sT\R\}$ of $\sT\R$ is a {\it Lie affgebroid} in the terminology of
\cite{GGU1, GGU2, GGU4}. Similarly, the affine hyperbundle $\wt{E}_1=E\ti\A_\R$ in $\wt{E}$ is an {\it
affgebroid} (so $\bE=\wt{E}_1\ti\R$ understood as the product in fibers is canonically a {\it special
affgebroid\ } in the terminology of \cite{GGU4}). The morphism (\ref{t-dep}) can be reduced then to
$$(\ze,{\zp}_{\A_\R}):\sT^\ast\wt{E}_1=\sT^\ast E\ti\sT^\ast\A_\R\ra\sT E^\ast\ti\A_\R\subset\sT(E^\ast\ti\R)\,.$$
Identifying $\A_\R$ with $\R$ in an obvious way, we obtain a {\it morphism of double affine bundles}
\cite{GGU4}
\be\label{td1}\bar{\ze}=(\ze,\bar{\zp}_\R):\sT^\ast({E}\ti\R)=\sT^\ast E\ti\sT^\ast\R\ra\sT E^\ast\ti\sT\R=\sT(E^\ast\ti\R)\,,\ee
where $\bar{\zp}_\R:\sT^\ast\R\ra \sT\R$ is defined by $\bar{\zp}_\R(t,s)=(t,1)\in\sT\R$.

Here, we view $\bar{E}=E\ti\R$ canonically as a vector bundle $\bar{\zt}:\bar{E}=E\ti\R\ra M\ti\R$ over
$M\ti\R$ (the pull-back bundle of $E$ with respect to the projection $M\ti\R\ra M$) and $E^\ast\ti\R$ as its
dual $\bar{E}^\ast$. The time-dependent analog of the diagram (\ref{diag}) defining the Tulczyjew
differential, for the time-dependent Lagrangian $L:E\ti\R\ra\R$ reads
\be\label{diag1}\xymatrix{
\sT^\ast (E\ti\R)\ar[rr]^{\bar{\ze}}  && \sT (E^\ast\ti\R)\ar[d]^{\zt_{({E}^\ast\ti\R)}} \\
E\ti\R\ar[rr]^{\bar{\zl}_L}\ar[u]^{\xd L}\ar@{.>}[rru]^{\bar{\Lambda}_{L}} &&
E^\ast\ti\R }.
\ee
Although there is a canonical identification $\A_\R\simeq\R$, the
use of $\A_\R$ explains the definition of holonomic vectors in
this case: since $\sT^{hol}(E\ti\A_\R)=\sT^{hol}E\ti\A_\R$, we
assume $\sT^{hol}(E\ti\R)=\sT^{hol}E\ti\R$. This is due to the
fact that the time-dependent picture is, in fact, an affgebroid
picture (see \cite{MMS,SMM,GGU4,IMPS,Ur1}).

In other words, $\bar{\zL}_L:E\ti\R\ra\sT E^\ast\ti \sT\R\simeq\sT(E^\ast\ti\R)$ and $\bar{\zl}_L:E\ti\R\ra
E^\ast\ti\R$ read
\be\label{nau}\bar{\zL}_L(e,t)=\left(\zL_{L^t}(e),(t,1)\right),\quad
\bar{\zl}_L(e,t)=(\zl_{L^t}(e),t)\,,\ee where we put $L^t(e)=L(e,t)$ and we canonically identified $\sT\R$
with $\R\ti\R$. If now $\zg$ is a curve in $E$, then the nonautonomous Euler-Lagrange equation reads
\be\label{ELna}(E_L^{na}):\qquad \st(\bar{\zl}_L\circ\bar{\gamma})=\bar{\Lambda}_L\circ\bar{\gamma},,\ee
where $\bar{\zg}(t)=(\zg(t),t)$ is a natural extension of $\zg$ to $E\ti\R$. The nonautonomous Euler-Lagrange
equation in coordinates takes formally the same form (\ref{EL2}), but now with $L$ depending on $t$.

\begin{ex}\label{e1}{\rm  There are many examples based on Lie algebroids, see for
instance \cite{CM,LMM,IMMS, Mar1, Mar3}.

\medskip\noindent {\bf (a)} For instance, for the canonical Lie algebroid and the corresponding morphism -- the inverse of the
Tulczyjew isomorphism \cite{Tu1}
$$\ze=\za_M^{-1}:\sT^\ast\sT M\ra\sT\sT^\ast M\,,$$
with $y^a=\dot{x}^a$, we get the traditional Euler Lagrange equations
$$\frac{\xd x^a}{\xd t}=\dot{x}^a, \quad
\frac{\xd}{\xd t}\left(\frac{\partial L}{\partial \dot{x}^a}\right)=\frac{\partial L}{\partial x^a}\,.$$

\medskip\noindent {\bf (b)} For a Lie algebroid which is just a Lie algebra with structure constants $c^k_{ij}$ with respect to a chosen
basis, we get the Euler-Poincar\'e equations
$$\frac{\xd}{\xd t}\left(\frac{\partial L}{\partial y^j}\right)= c^k_{ij} y^i\frac{\partial
L}{\partial y^k}\,.
$$

\medskip\noindent {\bf (c)} True Lie algebroid examples are  usually obtained as reductions of standard Lagrangian systems on tangent
bundles, like the reduction of the rigid body to a dynamics on $so(3,\R)$. Another example of this kind is a
homogeneous sphere of radius $r>0$, mass $m$, and inertia $k^2$ about any axis, moving on a horizontal table
without friction (thus, is the table rotating or not makes no difference). In an obvious way, the system lives
in fact on the Lie algebroid $\zt:\sT\R^2\ti so(3,\R)\ra\R^2$ with product Lie algebroid structure. In
standard coordinates the algebroid morphism
$$\ze:\sT^\ast\left(\sT\R^2\ti so(3,\R)\right)\ra\sT\left(\sT^\ast\R^2\ti so(3,\R)^\ast\right)$$ reads:
\bea\label{ex1}&\ze\left(x,y,\dot x,\dot y,\zw,p_x,p_y,p_{\dot x},p_{\dot y},p_\zw\right)=\\
&\left(x,y,p_{\dot x},p_{\dot y},p_\zw,\dot x,\dot
y,p_x,p_y,\zw_3p_{\zw_2}-\zw_2p_{\zw_3},\zw_1p_{\zw_3}-\zw_3p_{\zw_1},\zw_2p_{\zw_1}-\zw_1p_{\zw_2}\right)\,.
\nn\eea The pure kinetic Lagrangian
$$L=\frac{1}{2}m\left(\dot x^2+\dot y^2+k^2\left(\zw_1^2+\zw_2^2+\zw_3^2\right)\right)$$
induces the "free" dynamics
$$\frac{\xd}{\xd t}(m\dot x)=0,\quad \frac{\xd}{\xd t}(m\dot y)=0,\quad \frac{\xd}{\xd t}(mk^2\zw)=0\,.$$
Later we will add nonholonomic constraints to this picture.}
\end{ex}

\medskip
The above examples are associated with Lie algebroids, but some "nonholonomic constraints" on Lie algebroids
may lead to Lagrangian systems on quasi-Lie algebroids. This is related to quasi-Poisson brackets associated
with nonholonomic constraints \cite{Mr,SM}.

\begin{ex}{\bf (Algebroid of linear constraints)} {\rm Consider an algebroid structure on a vector bundle $E$ equipped
with a Riemannian metric $\la\cdot,\cdot\ran_E$ and a vector subbundle $C$ of $E$. Let $P:E\ra C$ be the
orthogonal projection. We can choose a local basis of orthonormal sections $(e_i)=(e_\za,e_A)$ of $E$ such
that $(e_\za)$ is a basis of local sections of $C$. According to the {\it d'Alembert principle} $\zd
L(\st(\zg)(t))\in C^0$, where $C^0\subset E^\ast$ is the annihilator of $C$, which in our case  (cf.
(\ref{EL2})) takes the form
$$\left(\frac{\xd}{\xd t}\left(\frac{\partial L}{\partial y^i}\right)- c^k_{\za i}(x) y^\za\frac{\partial
L}{\partial y^k} - \zs^a_i(x)\frac{\partial L}{\partial x^a}\right) e_i^*=\zm_A(x)e_A^*
$$
for certain functions $\zm_A$, the constrained dynamics is locally written as
\be y^A=0,\quad \frac{\xd x^a}{\xd t}=\zr^a_\za(x)y^\za, \quad
\frac{\xd}{\xd t}\left(\frac{\partial L}{\partial y^\zb}\right)- c^k_{\za\zb}(x) y^\za\frac{\partial
L}{\partial y^k} - \zs^a_\zb(x)\frac{\partial L}{\partial x^a}=0\,.\label{EL3}\ee  If we deal with a
Lagrangian of "mechanical type"
$$L=\frac{1}{2}(y^i)^2-V(x)\,,$$
then $\frac{\partial L}{\partial y^A}=y^A=0$ and the equations (\ref{EL3}) reduce to
$$y^A=0,\quad \frac{\xd x^a}{\xd t}=\zr^a_\za(x)y^\za, \quad
\frac{\xd}{\xd t}\left(\frac{\partial L}{\partial y^\zb}\right)- c^\zg_{\za\zb}(x) y^\za\frac{\partial
L}{\partial y^\zg} - \zs^a_\zb(x)\frac{\partial L}{\partial x^a}=0\,,$$ that can be viewed as the
Euler-Lagrange equations of the algebroid associated with the orthogonal projection of the tensor $\zP_\ze$
onto $C^\ast$ according to the orthogonal decomposition $E^\ast=C^0\oplus C^\ast$. Of course, even when $E$ is
a Lie algebroid, if $C$ is not a Lie subalgebroid, the projected tensor is not a Poisson tensor and we deal
with mechanics on a general algebroid, in fact a quasi-Lie algebroid in this case, since the projected Poisson
tensor remains skew-symmetric. }
\end{ex}

\section{Variational calculus}
For a general algebroid structure $\ze$ on the vector bundle $\zt:E\ra M$ and a smooth Lagrangian function
$L:E\ra\R$ we will define a version of a variational calculus as follows. Our (infinite-dimensional) manifold
$\cM$ will be the space of all $C^1$-paths $\zg:[t_0,t_1]\ra E$ in $E$. Of course, like in the standard
variational calculus, by curves through the path $\zg\in\cM$ we mean $C^1$-maps
$$h:[t_0,t_1]\ti\R\ni(t,s)\mapsto h(t,s)\in E$$
such that $h(t,0)=\zg(t)$. Thus, the tangent space $\sT_\zg\cM$ -- the space of all possible variations of
$\zg$ -- is represented by $\frac{\pa h}{\pa s}(t,0)$, i.e.  by continuous paths $\zd\zg:[t_0,t_1]\ra\sT E$
covering $\zg$ -- vector fields along $\zg$. The admissible paths form a subset $\cN$ which is a submanifold
in $\cM$ in a natural sense, since a path $\zg$ is admissible if and only if $\st(\zg)\subset \sT^{hol}E$. As
easily seen (see also \cite{Mar3}), a vector field $\zd\zg:[t_0,t_1]\ra E$ along an admissible path $\zg$
belongs to $\sT_\zg\cN$ if and only if $\zk_E\circ\st(\zd\zg)$ is tangent to $\sT^{hol}E$, i.e.
\be\label{tan}\zk_E(\st(\zd\zg)(t))\in\sT\sT^{hol}E\subset\sT\sT E\,,\ee
where $\zk_E:\sT\sT E\ra\sT\sT E$ is the canonical flip.

Note that we use here `infinite-dimensional manifold' structures in a very intuitive sense. However, we could
have put rigorously a Banach manifold structure on $\cM$, $\cN$, etc, similarly as it has been done in
\cite{Mar3}. On the other hand, because the Implicit Function Theorem will be not used, a less formal language
is completely satisfactory for our purposes, so we will skip technical complications associated with the
Banach manifold setting.

The Lagrangian $L$ defines a differentiable function (action functional) $W_L:\cM\ra\R$ by
\be\label{action}W_L(\zg)=\int_{t_0}^{t_1}L(\zg(t))\xd t\,.
\ee
Completely classically, the differential of the action $\xd W_L(\zg)$, paired with the tangent vector
$\zd\zg$, gives
\be\label{formula}\la\zd\zg,\xd W_L(\zg)\ran=\int_{t_0}^{t_1}\la\zd\zg(t),\xd
L(\zg(t))\ran\xd t.
\ee
Now, we will make use of the algebroid structure on $E$ and we will reduce the differential $\xd W_L$ to a
distribution $\cD$ over the submanifold $\cN$ in $\cM$ consisting of admissible paths. For an admissible path
$\zg:[t_0,t_1]\ra E$, the space ${\cD}(\zg)\subset\sT_\zg\cM$ of this distribution is exactly the space of
admissible variations (virtual displacements) $\zd_\zz\zg$ as they were defined in (\ref{tlift}), i.e.
 \be\label{v1}{\cD}(\zg)=\{\zd_\zz\zg: \zz\in \sV({\zg})\}\,.
 \ee
In this sense, the space $\sV(\zg)$ of vertical variations, which is geometrically well-understood as the
space of sections of the vertical bundle $\sV E$ along $\zg$, is a model space for the space $\cD(\zg)$ of
admissible (mechanical) variations which does not have so nice geometrical description in general. The reader
can easily check that in the case of the canonical Lie algebroid $E=\sT M$ the admissible variations we have
just introduced coincide with variations of tangent prolongations of paths in $M$ (with not fixed end-points
yet), as they are understood in Classical Mechanics. The geometrical meaning of these variations is usually
not understood being hidden behind the `obvious' Lie algebroid structure on $\sT M$.

Let us consider now the differential $\xd W_L$ being restricted to $\cD$. Our aim is to
show its special realization, very similar to the one present in the standard
variational calculus of Analytical Mechanics. Of special interest are variations
$\zd_\zz\zg$ coming from the set $\sV({\zg})_0$ of paths $\zz$ that vanish at the
end-points, $\zz(t_0)=0$, $\zz(t_1)=0$. They form a submanifold $\cD_0$ of $\cD$ and
analogs of the standard Euler-Lagrange equations are obtained as equations for critical
points of $(\xd W_L)_{\mid\cD_0}$, i.e. for such $\zg\in\cN$ that $\xd W_L(\zg)$ vanish
on $\cD_0(\zg)$. Note however, that in contrast with what has been done in \cite{Mar3},
being interested in the infinitesimal picture only, we do not care about global
homotopies inside the manifold of admissible paths. In fact, our distribution is not
tangent to $\cN$ in general, so even "infinitesimal homotopies" go outside $\cN$ in the
case of a general algebroid. This is due to the following observation.
\begin{theo} The distribution $\cD$ is tangent to the submanifold $\cN$ of admissible paths if and only if the
right and the left anchor coincide, $\zr=\zs$, and they induce a homomorphism of brackets:
\be\label{D}\zr([X,Y]_\ze)=[\zr(X),\zr(Y)]_{vf}\,,\ee
where $[\cdot,\cdot]_{vf}$ is the bracket of vector fields. In particular, $\cD\subset\sT\cN$ if $(E,\ze)$ is
a Lie algebroid.
\end{theo}
\begin{proof} It is a matter of easy calculations to show that, according to (\ref{tlift1a}),
the vector field $\zd_\zz\zg$ along $\zg(t)=(x(t),y(t))$ satisfies (\ref{tan}) if and only if
$$\frac{\xd f^j}{\xd t}(t)\left(\zs^b_j-\zr^b_j\right)(x(t))+f^j(t)y^i(t)\left(\frac{\pa\zs^b_j}{\pa
x^a}\zr^a_i-\frac{\pa\zr^b_i}{\pa x^a}\zs^a_j-c^k_{ij}\zr^b_k\right)(x(t))=0\,.$$ Since the above should be
satisfied for any admissible $\zg$ and for any given $x(t)=x(t_0)$, we can take $f^j(t_0)$, $\frac{\xd
f^j}{\xd t}(t_0)$ and $y(t_0)$ arbitrary. Hence we get $\zr=\zs$ and
$$\frac{\pa\zs^b_j}{\pa
x^a}\zr^a_i-\frac{\pa\zr^b_i}{\pa x^a}\zs^a_j-c^k_{ij}\zr^b_k=\frac{\pa\zr^b_j}{\pa
x^a}\zr^a_i-\frac{\pa\zr^b_i}{\pa x^a}\zr^a_j-c^k_{ij}\zr^b_k=0\,.$$ The latter can be rewritten in the form
$\zr([e_i,e_j]_\ze)=[\zr(e_i),\zr(e_j)]_{vf}$, whence (\ref{D}).
\end{proof}

\medskip\noindent
{\bf Remark.} One develops often a variational calculus introducing homotopies as "paths in path spaces"
satisfying certain boundary conditions -- this is exactly how the variational calculus on Lie algebroids has
been developed in \cite{Mar3}. However, this approach is much more restrictive when passing to constraints.
Let us only mention the existence of singular paths in the theory of linear nonholonomic constraints. In this
case no real variation of a singular path is possible, so the differential calculus does not make sense any
longer. On the other hand, the standard Euler-Lagrange equations are obtained as critical points of the action
- so in fact only "infinitesimal homotopies", i.e. admissible variations are used. In the Lie algebroid case,
the admissible homotopies can be taken as integral curves of variations. M.~Crainic and R.~L.~Fernandes have
related homotopies of admissible paths to flows of the complete lifts of time-dependent sections of the Lie
algebroid in their work \cite{CF} on integration of Lie algebroids. They did not mention the variational
calculus, but this integration is actually finding a manifold $G(E)$ (Lie groupoid) that allows to represent
the variational calculi on the Lie algebroid $E$ as reductions on standard variational calculus on $\sT G(E)$.
Let us also point out that, contrary to the approaches by M.~Crainic -- R.~L.~Fernandes and E.~Mart\'{\i}nez
\cite{CF,Mar3}, we work in full generality and we do not assume at the beginning that admissible variations
come from vertical variations vanishing at the end-points.

\bigskip Since calculating $\xd W_L(\zg)$ on $\cD$ according to (\ref{formula}), we can divide our path
into a finite number of smaller parts if needed, we can assume for simplicity that the
path $\zg$ lies in a single coordinate chart $(x^a,y^i)$, so we can write
$\zg(t)=(x^a(t),y^i(t))$. That our path is admissible means now that
\be\label{adm1}\frac{\xd x^a}{\xd
t}(t)=\zr^a_i(\ul{\zg}(t))y^i(t)\,.\ee For an admissible variation $\zd_\zz\zg$, with
$\zz(t)=f^i(t)e_i(\ul{\zg}(t))$, we have then
\beas\la\zd_\zz\zg(t),\xd
L(\zg(t))\ran&=&\Big[f^k(t)\cdot\zs^a_k(\ul{\zg}(t))\cdot\frac{\pa L}{\pa x^a}(\zg(t))+
\Big(y^i(t)\cdot c^j_{ik}(\ug(t))\cdot f^k(t)+ \frac{\xd f^j}{\xd t}(t)
\Big)\cdot\frac{\pa L}{\pa y^j}(\zg(t))\Big]\\
&=&f^k(t)\left(\zs^a_j(\ug(t))\cdot\frac{\pa L}{\pa x^a}(\zg(t))+ y^i(t)\cdot
c^k_{ij}(\ug(t))\cdot\frac{\pa L}{\pa y^k}(\zg(t))-\frac{\xd}{\xd t}\frac{\pa L}{\pa
y^k}(\zg(t))\right)+\\
&&\frac{\xd}{\xd t}\left(f^j(t)\frac{\pa L}{\pa y^j}(\zg(t))\right)\,.
\eeas

Writing $\zl_L:E\ra E^\ast$, $\zl_L(x,y)=\frac{\pa L}{\pa y^j}(x,y)e^j_\ast$, for the
vertical derivative (Legendre map) associated with $L$, and the {\it variation of the
Lagrangian} along $\zg$:
\be\label{dL}
\zd L\big(\st(\zg)(t)\big)=\left(\zs^a_j(\ug(t))\cdot\frac{\pa L}{\pa x^a}(\zg(t))+
y^i(t)\cdot c^k_{ij}(\ug(t))\cdot\frac{\pa L}{\pa y^k}(\zg(t))-\frac{\xd}{\xd
t}\frac{\pa L}{\pa y^j}(\zg(t))\right)e^j_\ast\,,
\ee
we can write \be\label{fff} \la\zd_\zz\zg(t),\xd L(\zg(t))\ran=\frac{\xd}{\xd t}\la
\zz_E(t),\zl_L(\zg(t))\ran+ \la \zz_E(t),\,\zd L\big(\st(\zg)(t)\big)\ran\,.
\ee
According to (\ref{F1.4a}), it is clear that $\zd L(\st(\zg)(t))=0$ if and only if the
image of the path $\xd L(\zg(t))$ under $\ze$ is admissible in $\sT E^\ast$, i.e. if
and only if $\zg$ satisfies the Euler-Lagrange equations (\ref{EL2}).

In a more explicit form the variation of the Lagrangian can be viewed as a map $\zd
L:\sT^{hol}E\ra E^\ast$ which in coordinates reads,
\bea\label{var}&\zd L(x,y,\dot{y})=\\ &\left(\zs^a_j(x)\frac{\pa L}{\pa
x^a}(x,y)+ y^i c^k_{ij}(x)\frac{\pa L}{\pa y^k}(x,y)-y^i\zr^a_i(x)\frac{\pa^2L}{\pa
x^a\pa y^j}(x,y)-\dot{y}^k\frac{\pa^2L}{\pa y^k\pa y^j}(x,y)\right)e^j_\ast\,.\nn
\eea
A geometrical description of the variation of the Lagrangian is as follows. If $v\in\sT
E$ is a holonomic vector, $v\in\sT^{hol}E$, then, as easily seen,
$$\zL_L\circ\zt_E\,,\sT\zl_L:\sT E\ra\sT E^\ast$$ are bundle maps over $\zl_L:E\ra
E^\ast$ and $\hat\zd L(v)=\zL_L(\zt_E(v))-\sT\zl_L(v)$ is a vertical vector in
$\sT_{\zt_E(v)}E^\ast$. As the vertical bundle $\sV E^\ast\subset\sT E^\ast$ is
canonically isomorphic to $E^\ast\oplus_M E^\ast$ by means of the vertical lift, we can
identify $\hat\zd L(v)$ with a vector $\zd L(v)=(\hat\zd L(v))_{E^\ast}$ from the fibre
of $E^\ast$ over $\zt(\zt_E(v))\in M$ which, expressed in coordinates, is exactly
(\ref{var}). In other words,
\be\label{vL}\zd L=\left((\zL_L\circ\zt_E-\sT\zl_L)_{\mid\sT^{hol}E}\right)_{E^\ast}=\left((\ze\circ\xd
L\circ\zt_E-\sT(\zt_{E^\ast}\circ\ze\circ\xd L))_{\mid\sT^{hol}E}\right)_{E^\ast}\,.\ee Using the obvious
pairing between $\sV E$ and $\sV E^\ast$ based on the fact that the fibers over $e$ and $e^*$, respectively,
are canonically dual spaces if $\zt(e)=\zp(e^*)$, we can write (\ref{fff}) equivalently in the form
\be\label{fff1}
\la\zd_\zz\zg(t),\xd L(\zg(t))\ran=\frac{\xd}{\xd t}\la \zz(t),\xd L(\zg(t))\ran+ \la
\zz(t),\,\hat\zd L\big(\st(\zg)(t)\big)\ran\,.
\ee
Integrating (\ref{fff}) (or (\ref{fff1})) we get
\bea\nn
\la\zd_\zz\zg,\xd W_L(\zg)\ran=\int_{t_0}^{t_1}\la\zd_\zz\zg(t),\xd L(\zg(t))\ran\xd t&=&\left.\la
\zz_E(t),\zl_L(\zg(t))\ran\right|^{t_1}_{t_0}+\int^{t_1}_{t_0}\la
\zz_E(t),\,\zd L\big(\st(\zg)(t)\big)\ran\,\xd t\\
&=&\left. \zz(L)(\zg(t))\right|^{t_1}_{t_0}+\int^{t_1}_{t_0}\la \zz(t),\,\wh\zd
L\big(\st(\zg)(t)\big)\ran\,\xd t \,.\label{total}
\eea
Now, if $\zz\in \sV({\zg})_0$, then $ \la \xd
W_L(\zg),\zd_\zz\zg\ran=\int^{t_1}_{t_0}\la \zz_E(t),\,\zd
L\big(\st(\zg)(t)\big)\ran\,\xd t$. If $\zz\in \sV({\zg})_0$, then $r(t)\zz(t)\in
\sV({\zg})_0$ for any function $r:[t_0,t_1]\ra\R$, so $\xd W_L(\zg)$ vanishes on
$\cD_0(\zg)$ if and only if $\zd L(\st(\zg))=0$.

We can summarize the above observations as follows.
\begin{theo}\label{t*} By means of the variational calculus for a general algebroid
one can define the {\it velocities-momenta} correspondence (Legendre map): $\zl_L:E\ra E^\ast$ and the
variation of the Lagrangian $\zd L:\sT^{hol}E\ra E^\ast$,  such that the derivative of the action functional
$\xd W_L(\zg)$ is represented by
$$ \la \xd W_L(\zg),\zd_\zz\zg\ran=\left.\la
\zz_E(t),\zl_L(\zg(t))\ran\right|^{t_1}_{t_0}+\int^{t_1}_{t_0}\la \zz_E(t),\,\zd
L\big(\st(\zg)(t)\big)\ran\,\xd t\,.
$$ Moreover, the formula (\ref{vL}) defines the Tulczyjew differential $\zL_L:E\ra\sT E^*$ associated with the
Lagrangian $L$. An admissible path $\zg(t)=(x(t),y(t))$ in $E$ satisfies $\zd
L(\st(\zg))=0$ if and only if $\xd W_L$ vanishes on $\cD_0(\zg)$ and if and only if
$\zg$ satisfies the Euler-Lagrange equations (\ref{EL2}).
\end{theo}\noindent
For a given admissible path $\zg:[t_0,t_1]\ra E$, the values $p(t_0)=\zl_L(\zg(t_0))$ and
$p(t_1)=\zl_L(\zg(t_1))$ represent the {\it initial} and the {\it final momenta}, and $\zh_\zg(t)=\zd
L(\st(\zg)(t))$ -- the {\it external force} that we have to apply to make the system moving along the path
$\zg$. A standard way to obtain the dynamics in Analytical Mechanics is to look for critical points of the
action functional with respect to admissible variations $\zd_\zz\zg$ which vanish at the end-points. In this
way we obtain the Euler-Lagrange equations (\ref{EL2}) for admissible curves in the form $\zd
L(\st(\zg)(t))=0$. In a more general setting, one can view the force defining equation
\be\label{force}
\zd L(\st(\zg)(t))=\zh_\zg(t)\ee as a differential equation for $\zg$ if the external force $\zh_\zg(t)$ is
given. In many cases this force is defined in path-independent way as a time-dependent field of forces
$F:E\ti\R\ra E^*$, $\zp(F(e,t))=\zt(e)$, i.e. $\zh_\zg(t)=F(\zg(t),t)$.

\medskip
There is no real difference when we admit time-dependent Lagrangians $L:E\ti\R\ra\R$,
so that the action reads
\be\label{action1}W_L(\zg)=\int_{t_0}^{t_1}L(\zg(t),t)\xd t\,.\ee
The formula (\ref{total}) just takes the form
\be\label{total1}
\la \xd W_L(\zg),\zd_\zz\zg\ran=\left.\left(f^j(t)\cdot\frac{\pa L}{\pa
y^j}(\zg(t),t)\right)\right|^{t_1}_{t_0}+\int^{t_1}_{t_0}\la \zz_E(t),\,\zd
L\big(\st(\zg)(t),t\big)\ran\,\xd t\,,
\ee
where
\be\label{dL1}\zd L\big(\st(\zg)(t),t\big)=\left(\zs^a_j(\ug(t))\cdot\frac{\pa L}{\pa
x^a}(\zg(t),t)+ y^i(t)\cdot c^k_{ij}(\ug(t))\cdot\frac{\pa L}{\pa
y^k}(\zg(t),t)-\frac{\xd}{\xd t}\frac{\pa L}{\pa y^j}(\zg(t),t)\right)e^j_\ast\,.\ee In
local coordinates,
\bea\label{var1}&\zd L(x,y,t,\dot{y})=\\ &\left(\zs^a_j(x)\frac{\pa L}{\pa
x^a}(x,y,t)+ y^i c^k_{ij}(x)\frac{\pa L}{\pa y^k}(x,y,t)-y^i\zr^a_i(x)\frac{\pa^2L}{\pa x^a\pa
y^j}(x,y,t)-\dot{y}^k\frac{\pa^2L}{\pa y^k\pa y^j}(x,y)-\frac{\pa^2L}{\pa t\pa
y^j}(x,y,t)\right)e^j_\ast\,.\nn
\eea
The geometrical picture is based on (\ref{diag1}). Now, $\zd L:\sT^{hol}E\ti\R\ra E^*$ is defined as the map
whose vertical lift is
\be\label{dl1}\hat\zd L=\textsf{v}_\zp\circ\zd L=
(\bar{\zL}_L\circ\zt_{\bar{E}}-\sT\bar{\zl}_L)_{\mid \sT^{hol}E\ti\R}=
(\bar{\ze}\circ\xd L\circ\zt_{\bar{E}}-\sT(\zt_{\bar{E}^\ast}\circ\bar{\ze}\circ\xd
 L))_{\mid \sT^{hol}E\ti\R}\ee
and the standard Euler-Lagrange equation for the time-dependent Lagrangian with a presence of external forces
takes the form $$\zd L(\st(\zg)(t),t)=\zh_\zg(t).$$ Again, the equation $\zd L(\st(\zg)(t),t)=0$ means that
the image of the path $\xd L(\zg(t),t)$ in $\sT^\ast(E\ti\R)$ under $\bar{\ze}$ is an admissible path (tangent
prolongation) in $\sT(E^\ast\ti\R)$.

\section{Constraints} In view of the just developed variational calculus on general algebroids we can introduce, in
principle, two types of constraints: the {\it configuration constraints} which are put in the "bundle of
velocities" $E$, i.e. constrains for paths in $\cN$, and the {\it virtual displacement constraints} put for
variations, i.e. for fibers of the admissible distribution $\cD$. As the admissible variations are also
related to paths in $E$, the latter constraints can be defined also via constraints in $E$ that often leads to
misunderstandings. In this way, a constrained submanifold in $E$ (classically in $E=\sT M$) is sometimes
referred to as a {\it nonholonomic constraint}. Note however that in general, speaking on a submanifold (in
general -- subset) of $E$ as of a constraint does not make much sense before we decide how the constrained
submanifold produces true constraints in the variational calculus. To put some order in the subject, we will
start with describing our understanding of constraints in the variational calculus for general algebroid that
will motivate a description of constraints in the pure geometric setting.
\begin{Definition}{\rm A {\it constraint} in the variational calculus for a general
algebroid is a subset $\cC$ of the bundle $\cD$. The corresponding (dynamical) {\it
configuration constraint} is the subset $\cC_\cN$ obtained from $\cC$ by the projection
to $\cN$. The {\it constrained variational calculus} is the study of the differential
of the action functional $\xd W_L$ restricted to $\cC$, or $\cC_0=\cC\bigcap\cD_0$. }
\end{Definition}

\medskip\noindent
It seems that the true variational constraints in physics strongly depend on the actual
system we work with. In theory however, the variational constraints are often derived
from {geometric constraints} of different types in a more or less canonical way. A {\it
geometric constraint} will be understood as a submanifold (more generally - a subset)
$S$ in $E$. Of course, as we have already mentioned (see also \cite{Tu2}), the
submanifold (subset) $S\subset E$ does not define a true variational constraint without
additional specifications. There are at least two geometrically justified ways of
deriving variational constraints out of $S$. According to the tradition (see the review
article \cite{CLMM1}), we will refer to them, respectively, as to {\it vakonomic} and
{\it nonholonomic} constraints. In the vakonomic case we accept only admissible
variations (virtual displacements) which are tangent to the constraint, while in the
nonholonomic case we admit only vertical variations which are tangent to the
constraint, i.e. which belong to $\sV(S)=\sT S\bigcap\sV E$.

\begin{Definition}\

{\rm \begin{itemize}\item The {\it vakonomic constraint} associated with $S\subset E$
is the variational constraint $\cC^{vk}(S)$ consisting of these admissible variations
$\zd_\zz\zg$ which are tangent to $S$, i.e. $\zd_\zz\zg(t)\in\sT S$. In particular, the
admissible path $\zg$ lies in $S$. \item The {\it nonholonomic constraint} associated
with $S\subset E$ is the variational constraint $\cC^{nh}(S)$ consisting of admissible
variations $\zd_\zz\zg$ associated with vertical variations $\zz$ which are tangent to
$S$. In other words, $\zz(t)\in\sT S$ (thus, $\zz(t)\in\sV(S)$). In particular, the
admissible path $\zg$ lies in $S$. \item A geometric constraint $S\subset E$ we call
{\it holonomic}, if the nonholonomic variational constraint associated with $S$ implies
the vakonomic constraint, i.e. $\cC^{nh}(S)\subset\cC^{vk}(S)$.
\end{itemize}}
\end{Definition}
\noindent Note that the variational constraints associated with $S$ can be very small
or even empty, e.g. when there are no admissible paths in $S$. To avoid pathologies
like that, certain additional {integrability conditions} can be introduced. A natural
{\it integrability condition} we will use is $\zr(S)\subset\sT S_M$, where
$S_M=\zt(S)$. It is assumed in the sequel that the geometric constraints are
integrable.

\medskip\noindent {\bf Remark.} We should stress here the obvious fact that $\sT S$ is well defined
in a general setting even if $S$ is not a submanifold of $E$, since it makes sense to
speak about smooth curves in $E$ with values in $S$. Note that, just by definition, for
the vakonomic constraint only the restriction of the Lagrangian function $L$ to $S$
plays the role in the variational problem. The latter is not the case for nonholonomic
constraints, except for the holonomic case. We can say that holonomic constraints are
those nonholonomic constraints for which only the restrictions of the Lagrange
functions to $S$ play the role in the corresponding variational problems. One can
easily derive from the form of the lift (\ref{tlift1a}) that a linear (integrable)
constraint $S$, i.e. a vector subbundle $S\subset E$, is holonomic if and only if the
algebroid bracket $[\cdot,\cdot]_\ze$ is closed on sections of $S$, i.e. $S$ a
subalgebroid in $E$.

\medskip\noindent
{\bf 1. Vakonomic constraints -- variational approach.} The variational problem depends
now on studying the differential of the action functional on $\cC^{vk}(S)$. A naive but
instructive approach is that the corresponding constrained Euler-Lagrange equations
describe admissible paths $\zg$ in $S$ which are {\it critical points of $W_L$ relative
to the generalized distribution $\cC^{vk}_0(S)=\cC^{vk}(S)\bigcap\cD_0$}, i.e. such
that $\xd W_L(\zg)$ vanishes on all $\zd_\zz\zg\in\cC^{vk}_0(S)(\zg)$:
\be\label{vp}\la \xd
W_L(\zg),\zd_\zz\zg\ran=\int_{t_0}^{t_1}\la\xd L(\zg(t)),\zd_\zz\zg(t)\ran\xd
t=\int^{t_1}_{t_0}\la \zz_E(t),\,\zd L\big(\st(\zg)(t)\big)\ran\,\xd t=0\ee for all
vertical vector fields $\zz$ along ${\zg}$, with $\zz(t_0)=0$, $\zz(t_1)=0$, and such
that $\zd_\zz\zg$ is tangent to $S$. Of course, it is hard to decide how large is
$\cC^{vk}_0(S)$. This is related to the difficult questions of the existence of
singular or abnormal paths, etc., which cannot be solved in the whole generality.
Leaving these questions aside, we will reduce ourselves to natural and geometric
sufficient conditions ensuring that a given admissible path satisfies (\ref{vp}).
Namely, let us observe that if $\zF$ is a function vanishing on $S$, then, as
$\zd_\zz\zg$ is tangent to $S$,
$$\la\xd \zF(\zg(t)),\zd_\zz\zg(t)\ran=0\,.$$
Thus, if
\be\label{vc}\zd(L-\zm_k\zF^k)(\st(\zg)(t),t)=0\,,\quad \zg(t)\in S\,,
\ee
for certain $\zm_k:[t_0,t_1]\ra\R$ and for certain functions $\zF^k$ vanishing on $S$
(e.g. defining $S$ according to the Implicit Function Theorem), then, according to
(\ref{vp}) applied to $L:=L-\zm_k\zF^k$,
\beas &\la \xd W_L(\zg),\zd_\zz\zg\ran=\int_{t_0}^{t_1}\left\la\xd
L(\zg(t)),\zd_\zz\zg(t)\right\ran\xd t\\&=\int_{t_0}^{t_1}\left\la\left(\xd
L-\zm_k(t)\xd\zF^k\right)\left(\zg(t)\right),\zd_\zz\zg(t)\right\ran\xd
t=\int^{t_1}_{t_0}\left\la \zz_E(t),\,\zd\left(
L-\zm_k\zF^k\right)\big(\st(\zg)(t),t\big)\right\ran\,\xd t=0\,,\eeas so (\ref{vp}) is
satisfied. Such $\zg$ we will call a {\it normal solution} of the vakonomic variational
problem associated with $S\subset E$. In the above procedure we can take as well a
time-dependent Lagrangian $L$ satisfying (\ref{vc}). The latter does not depend
directly on how big is $\cC^{vk}_0(S)$ and it simply means that the image of the path
$\xd\left(L-\zm_k\zF^k\right)(\zg(t),t)$ in $\sT^\ast(E\ti\R)$ under $\bar{\ze}$ is an
admissible path (tangent prolongation) in $\sT(E^\ast\ti\R)$. Motivated by the
tradition we will regard the equation (\ref{vc}) as {\it vakonomically constrained
Euler-Lagrange equation}. There is a clear analog of the above procedure also for
time-dependent constraints. The non-autonomous vakonomic Euler-Lagrange equation takes
in coordinates the form
\bea\label{ELvc}
&\zF^k(x,y)=0\,,\quad
\frac{\xd x^a}{\xd t}=\zr^a_k(x)y^k\,,\\
&\frac{\xd}{\xd t}\frac{\partial L}{\partial y^j}(x,y,t)-c^l_{ij}(x) y^i\frac{\partial L}{\partial y^l}(x,y,t)
- \zs^a_j(x)\frac{\partial L}{\partial x^a}(x,y,t)
=\label{ELvc1}\\
&\dot{\zm}_k(t)\frac{\partial \zF^k}{\partial y^j}(x,y)+\zm_k(t)\left(\frac{\xd}{\xd t}\frac{\partial
\zF^k}{\partial y^j}(x,y) -c^l_{ij}(x) y^i\frac{\partial \zF^k}{\partial y^l}(x,y) - \zs^a_j(x)\frac{\partial
\zF^k}{\partial x^a}(x,y)\right)\nn
\eea
and reduces to the classical one for the canonical Lie algebroid $E=\sT M$ (see e.g.
\cite{CLMM1}). In the above form, however, the vakonomic Euler-Lagrange equation is not
easily seen to depend only on the restriction of $L$ to $S$. Below we present a
geometric approach clarifying this question. On the other hand, this approach seems to
be more transparent not only at this point.

\medskip\noindent
{\bf 2. Vakonomic constraints -- geometric approach.} Of course, one can take
(\ref{ELvc}) as the Euler-Lagrange equation for the vakonomic constraints without
referring to the variational calculus. It has the advantage that we do not care about
possibly complicate constrained admissible variations. There is a nice geometric
interpretation of these equations. For simplicity we will reduce to the autonomous
case, so that $L$ does not depend on $t$. Let us recall first that, with any
submanifold $S$ in $E$ and any function $L:S\ra \R$ one can associate canonically a
lagrangian submanifold $S_L$ in $\sT^\ast E$ defined by
$$S_L=\{ \za_e\in\sT^\ast_eE: e\in S\text{\ and\ }\la\za_e,v_e\ran=\xd L(v_e)\text{\ for every\ }
v_e\in\sT_eS\}\,.$$ If $S=E$, then $S_L=\xd L(E)$, i.e. $S_L$ reduces to the image of
$\xd L$. We have the following.
\begin{theo} A curve $\zg:\R\ra E$ satisfies the vakonomic Euler-Lagrange equations (\ref{vc})
associated with an autonomous Lagrangian $L:E\ra\R$ if and only if it is a projection
to $S$ of a curve $\zg^\ast:\R\ra S_L$ whose image under $\ze:\sT^\ast E\ra\sT E^\ast$
is admissible (is a tangent prolongation of a curve in $E^\ast$). In particular, the
vakonomic Euler-Lagrange equations depend on the restriction of the Lagrangian to the
constraint only.
\end{theo}
\begin{proof} If a curve $\zg$ satisfies (\ref{vc}), then $\zg$ is
admissible and lies in $S$. Moreover, the curve $\zg^*(t)=\xd
L(\zg(t))-\zm_k(t)\xd\zF^k(\zg(t))$ in $\sT^\ast E$ lies in $S_L$, projects on $\zg$
and is mapped trough $\ze$ to an admissible curve.

Conversely, if a curve $\zg^*(t)$ has the above properties, then there are $\zm_k(t)$
such that $\zg^*(t)=\xd L(\zg(t))-\zm_k(t)\xd\zF^k(\zg(t))$. Since admissibility of
$\ze(\zg^*(t))$ is is equivalent to $\zd\left(\xd L-\zm_k(t)\xd\zF^k\right)(\zg(t))=0$
(Theorem \ref{t*}), the theorem follows.
\end{proof}

\medskip\noindent One can also think that the vakonomic Euler-Lagrange equations are not
equations on curves in $E$ but on curves in $S_L$. Then, we can just consider the
projections of the solutions onto $E$.

The corresponding diagram is the following
\be\label{vkc} \xymatrix{ \sT^\ast E\supset S_L
\ar[rr]^{\ze}&& \sT E^\ast\ar[d]^{\zt_{E^\ast}}\\ S\ar @{-|>}[u]^{r_L}\ar@{.|>}[rru]_{{\zL}_{\, L}} \ar
@{-|>}[rr]_{{\zl}_{\,L}} &&E^\ast}
\ee
where $r_L$ is the relation which is the inverse of the projection $(\zp_E)_{\mid
S_L}:S_L\ra S$ and $\zL_L=\ze\circ r_L$. Like in the non-constrained case, a curve
$\zg$ in $S$ satisfies the vakonomic E-L equation if it is related {\it via}
${\zL}_{\,L}$ to an admissible curve in $\sT E^\ast$.

From the above it should be clear that the phase space for the vakonomic constraint $S$
is $\zt_{E^\ast}(\ze(\zp^{-1}(S)))$ and the phase dynamic associated with the
Lagrangian $L$ is $\ze(S_L)$. There is an obvious version of the above picture in the
non-autonomous case.
\begin{ex} {\bf (Pontryagin's Maximum Principle)} {\rm For an algebroid $(E,\ze)$ over $M$ consider the product
algebroid $E_U=E\ti\sT U$. Considering an optimal control problem in which the manifold $U$ plays the role of
set of control parameters and  associated with: \begin{enumerate} \item an integrable constraint $S$ defined
by means of a $U$-dependent section $f:M\ti U\ra E$ of $E$ by $(e,v)\in S\ \Leftrightarrow
e=f(\zt(e),\zt_M(v))$ and \item a Lagrangian function $L:S\ra\R$ depending only on the base,
$L(e,v)=L(\zt(e),\zt_M(v))$.
\end{enumerate}
In local coordinates $(x^a,u^\za,y^i,\dot u^\zb)$ in $E_U$ and the adapted coordinates $(x,u,y,\dot
u,p_x,p_u,\zx,\zp)$ in $\sT^\ast E_U$, the product algebroid morphism $\ze_U=(\ze,\ze_M)$ reads
$$\ze_U\left(x,u,y,\dot u,p_x,p_u,\zx,\zp\right)=\left(x,u,\zx,\zp,\zr^b_k(x)y^k, \dot u,c^k_{ij}(x)
y^i\zx_k + \zs^a_j(x)p_{x^a},p_u\right)\,.$$ The Lagrangian submanifold $S_L^\ast\subset\sT^\ast E_U$ consists
of points
$$\left(x,u,f(x,u),\dot u,\left(\frac{\pa L}{\pa x}-\zx\cdot\frac{\pa f}{\pa x}\right)(x,u),
\left(\frac{\pa L}{\pa u}-\zx\cdot\frac{\pa f}{\pa u}\right)(x,u),\zx,0\right)\,,$$ so the phase (implicit)
dynamics is given by $\ze_U(S_L^\ast)$, which is the set of points
$$\left(x,u,\zx,0\,,\zr^b_k(x)f^k(x,u)\,,\dot u,c^k_{ij}(x)
f^i(x,u)\zx_k + \zs^a_j(x)\left(\frac{\pa L}{\pa x^a}-\zx\cdot\frac{\pa f}{\pa x^a}\right)(x,u)\,,
\left(\frac{\pa L}{\pa u}-\zx\cdot\frac{\pa f}{\pa u}\right)(x,u)\right)\,,$$ and the vakonomic Euler-Lagrange
equations read
\bea \label{pont1}&\frac{\xd x^b}{\xd t}=\zr^b_k(x)f^k(x,u),\\
\label{pont2}&\frac{\xd \zx_j}{\xd t}=c^k_{ij}(x) f^i(x,u)\zx_k + \zs^a_j(x)\left(\frac{\pa L}{\pa
x^a}-\zx_i\frac{\pa f^i}{\pa x^a}\right)(x,u)\,,\\\label{pont3}& \left(\frac{\pa L}{\pa u}-\zx_i\frac{\pa
f^i}{\pa u}\right)(x,u)=0\,.
\eea
The equations (\ref{pont1}) and (\ref{pont2}) describe the phase dynamics on $E^\ast\ti U$  associated with
the Hamiltonian $H(x,u,\zx)=f^i(x,u)\zx_i-L(x,u)$ via the tensor $\zP_\ze$ -- trivially extended from $E^\ast$
to $E^\ast\ti U$. The equation (\ref{pont3}) in turn, is the equation for critical points of this Hamiltonian
with respect to the control variable $u$. In the classical case $E=\sT M$, the equations (\ref{pont2}) and
(\ref{pont3}) read
$$\frac{\xd p_a}{\xd t}=\left(\frac{\pa L}{\pa x^a}-p_b\frac{\pa
f^b}{\pa x^a}\right)(x,u)\,,\quad \left(\frac{\pa L}{\pa u^\za}-p_b\frac{\pa f^b}{\pa
u^\za}\right)(x,u)=0\,.$$ We recognize the Pontryagin's Maximum Principle in its normal
differential form. For Lie algebroids, this principle was first proposed in
\cite{Mar4}.}
\end{ex}

\medskip\noindent
{\bf 3. Nonholonomic constraints -- variational approach.} A naive but instructive approach is to assume in
this case that the constrained Euler-Lagrange equations describe admissible paths $\zg$ in $S$ which are {\it
critical points of $W_L$ relative to the generalized distribution $\cC^{nh}_0(S)$}, i.e. such that (\ref{vp})
is satisfied for all $\zd_\zz\zg\in \cC^{nh}_0(S)$. Again, we will not discuss the problem how large is
$\cC^{nh}_0(S)$. Recall that $\zd_\zz\zg\in \cC^{nh}(S)$ means that $\zz(t)\in \sV(S)$, where $\sV(S)=\sT
S\bigcap\sV E$ is the vertical part of $\sT S$. If $S$ is a submanifold and $\sV(S)$ has constant rank, then
the annihilator $(\sV(S))^0\subset\sT^\ast E_{\mid S}$ is a vector subbundle (over $S$) in $\sT^\ast E$. In
this case the quotient bundle $\sT^\ast E_{\mid S}/(\sV(S))^0$ is canonically isomorphic to the bundle
$\sV^\ast(S)$ -- dual to $\sV(S)$. The latter, viewed as a subbundle in $pr_1:E\oplus_M E\ra E$ in an obvious
way, is called the {\it bundle of virtual displacements} in \cite[Section 8]{CLMM}. Of course, $(\sV(S))^0$
can be viewed in a similar way as a subbundle in $E\oplus_M E^\ast\ra E$. In this interpretation, which we
will generally use in the sequel, $(\sV(S))^0_e\subset E^\ast_{\zt(e)}$, $e\in S$, is the annihilator of
$(\sV(S))_e\subset E_{\zt(e)}$ and $(\sV^\ast(S))_e=E^\ast_{\zt(e)}/(\sV(S))^0_e$.

It is obvious that (\ref{vp}) is satisfied for all $\zd_\zz\zg\in \cC^{nh}_0(S)$ if
(and not only if, in general)
\be\label{nhv}\zd L(\st(\zg)(t))\in (\sV(S))^0_{\zg(t)}\,.\ee
This equation we will view as the {\it constrained nonholonomic
Euler-Lagrange equation}. Again, it is not exactly equivalent to
the variational principle in general as it gives only a sufficient
condition for a relative critical point of the action functional
$W_L$. In local coordinates, if $\zF^k$ are functions defining the
constraint $S$ {\it via} equations $\zF^k(x,y)=0$, then $\sV(S)^0$
is generated by $\frac{\pa\zF^k}{\pa y^i}$ at points of $S$, so
the constrained nonholonomic Euler-Lagrange equation reads
\bea\label{nhlc} &\zF^k(x,y)=0, \quad \frac{\xd x^a}{\xd t}=\zr^a_i(x)y^i\\&
\frac{\xd}{\xd t}\frac{\partial L}{\partial y^j}(x,y)- c^l_{ij}(x) y^i\frac{\partial L}{\partial y^l}(x,y)
-\zs^a_j(x)\frac{\partial L}{\partial x^a}(x,y)=\zm_k(t)\frac{\pa\zF^k}{\pa y^j}(x,y)\,. \label{nhlc4}
\eea
For $E=\sT M$ this is exactly the {\it Chetaev principle} and for $E$ being an
arbitrary Lie algebroid the equations (\ref{nhlc}), (\ref{nhlc4}) coincide with the
equations associated with nonlinear nonholonomic constraints considered in
\cite{CLMM,LMD}.
\begin{ex} {\bf (rolling ball)} {\rm Consider now the celebrated example of a ball rolling on a rotating
table (cf. \cite{BKMM,CLMM}), more precisely, of a homogeneous sphere of radius $r>0$, mass $m$, and inertia
about any axis $k^2$, moving without sliding on a horizontal table which rotates with constant angular
velocity $\zW$. Like in Example \ref{e1}, we can recognize that the system lives on the Lie algebroid
$\zt:\sT\R^2\ti so(3,\R)\ra\R^2$ with product Lie algebroid structure and is ruled by the pure kinetic
Lagrangian
$$L=\frac{1}{2}m\left(\dot x^2+\dot y^2+k^2\left(\zw_1^2+\zw_2^2+\zw_3^2\right)\right)\,,$$
this time however with the presence of nonholonomic constraints
\beas\zF^1(x,y,\dot x,\dot y,\zw)&=&\dot x-r\zw_2+\zW y=0\,,\\
\zF^2(x,y,\dot x,\dot y,\zw)&=&\dot y+r\zw_1-\zW y=0\,.
\eeas
According to (\ref{nhlc}) and (\ref{nhlc4}), we get the constrained nonholonomic Euler-Lagrange equation in
the form
\beas &\dot x-r\zw_2+\zW y=0,\quad \dot y+r\zw_1-\zW y=0,\quad
\frac{\xd}{\xd t}(m\dot x)=\zm_1,\quad \frac{\xd}{\xd t}(m\dot y)=\zm_2,\\
& \frac{\xd}{\xd t}(mk^2\zw_1)=r\zm_2,\quad \frac{\xd}{\xd t}(mk^2\zw_2)=-r\zm_1,\quad\frac{\xd}{\xd
t}(mk^2\zw_3)=0\,,
\eeas
that easily implies
\beas\ddot x+\frac{k^2\zW}{r^2+k^2}\dot y&=&0\,,\\
\ddot y-\frac{k^2\zW}{r^2+k^2}\dot x&=&0\,.\eeas }
\end{ex}

\medskip
If $S$ is a linear constraint, i.e. $S$ is a vector subbundle in $E$, then $(\sV(S))_e$
can be identified with $S_{\zt(e)}$ and $(\sV(S))_e^0$ with $S_{\zt(e)}^0\subset
E^\ast_{\zt(e)}$. In this case the constrained nonholonomic Euler-Lagrange equation
(\ref{nhv}) takes the form
\be\label{nhv*}\zd L(\st(\zg)(t))\in S^0_{\ul{\zg}(t)}\,,\ee
which is exactly the {\it d'Alembert's principle} of virtual work.
The d'Alembert's principle for Lie algebroids was first proposed
in \cite{CM}.

More generally, assume that $S=A$ is an affine constraint, i.e.
$A$ is an affine subbundle in $E$. Then, $\sV(A)_e$ can be
canonically identified with the fiber $\sv(A)_{\zt(e)}\subset
E_{\zt(e)}$ of a vector bundle $\sv(A)$ which serves as a model
vector bundle of $A$. Hence, $(\sV(S))_e^0$ can be identified with
$\sv(A)_{\zt(e)}^0\subset E_{\zt(e)}^0$ and the constrained
nonholonomic Euler-Lagrange equation reads
\be\label{nhv2}\zd L(\st(\zg)(t))\in (\sv(A))^0_{\ul{\zg}(t)}\,.\ee

\medskip\noindent
{\bf 4. Affine nonholonomic constraints -- geometric approach.} Let us assume that
$S=A$ is an affine subbundle in $E$ (over $A_M$) satisfying the integrability condition
$\zr(A)\subset\sT A_M$. In this case $(\sV(A))_e$ is constant along fibers of $A$ and
it coincides with the $\sv(A)_{\zt(e)}$. Let $\sv(A)^0$ be the annihilator of $\sv(A)$
which is a subbundle in $E^\ast$ (over $A_M$). Let $i_{\sv(A)}:\sv(A)\hookrightarrow E$
be the inclusion of $\sv(A)$ in $E$, let $i_{\sv(A)}^\ast:E^\ast_{\mid A_M}\ra
\sv(A)^\ast$ be the dual of $i_{\sv(A)}$, and let
$$\sT{i_{\sv(A)}^\ast}:\sT (E^\ast_{\mid A_M})\ra\sT \sv(A)^\ast$$
its tangent prolongation. According to the integrability condition $\zr(A)\subset\sT A_M$, the image
$\ze(\sT^\ast E_{\mid A})$ lies in $\sT(E^\ast_{\mid A_M})$ and the corresponding diagram is the following
\be\label{nhc} \xymatrix{ \sT^\ast E_{\mid A}\ar[rr]^{\ze}  && \sT (E^\ast_{\mid
A_M})\ar[rr]^{\sT{i_{\sv(A)}^\ast}}&&
\sT \sv(A)^\ast\ar[d]^{\zt_{\sv(A)^\ast}}\\
E\supset A\ar[rr]^{\zl_L}\ar[u]^{\xd
L}\ar@{.>}[rrrru]^{\zL_L^A}\ar[rr]^{\zl_{L}}\ar@/d3ex/@{.>}[rrrr]_{\zl_L^A} && E^\ast_{\mid
A_M}\ar[rr]^{i_{\sv(A)}^\ast} && \sv(A)^\ast}
\ee
The space $\sv(A)^\ast$ is the phase space for the nonholonomic constraint $A$ with $\zl_L^A:A\ra
\sv(A)^\ast$, $\zl_L^A=i_{\sv(A)}^\ast\circ\zl_L$, as the {\it constrained Legendre map}, and $\zL_L^A:A\ra\sT
{\sv(A)}^\ast$, with $\zL_L^A=\sT{i_{\sv(A)}^\ast}\circ\ze\circ\xd L$, serves as the {\it constrained
Tulczyjew differential}. The set $\zL_L^A(A)\subset\sT \sv(A)^\ast$ is the phase dynamics associated with the
Lagrangian $L$. The nonholonomic Euler-Lagrange equation is described as follows.
\begin{theo} A curve $\zg:\R\ra A$ satisfies the nonholonomic Euler-Lagrange equation $\zd L(\st(\zg)(t))\in
\sv(A)^0_{\ul{\zg}(t)}$ if and only if the curve $\zL_L^A(\zg(t))$ in $\sT \sv(A)^\ast$
is admissible (is the tangent prolongation of a curve in $\sv(A)^\ast$).
\end{theo}
\begin{proof}Consider local coordinates $(x^I)=(x^i,x^\zi)$ on a open set ${U}$ of $M$ such that $A_M$ is determined by
the constraint $x^\zi=0$. A local basis $\{e_a\}_{a=1,\dots,n-r}$ of sections of $\sv(A)$ together with a
section $e_0$ of $A$ we can extend to local sections of $E$ and complete them to a local basis of sections $\{
e_0, e_a, e_{\alpha}\}$ of the vector bundle $E$. Then, in coordinates
$(x^I,y^A)=(x^i,x^\zi,y^0,y^a,y^\alpha)$ adapted to this bases, the local equations defining the constrained
subbundle $A$ as an affine subbundle of $E$ over $A_M$ are $x^\zi=0$, $y^0=1$, $y^\alpha=0$, so points of $A$
have coordinates $(x^i,0,1,y^a,0)$. Note that integrability of the constraint $A$ means that $\zr_0^\zi(x)=0$
and $\zr_a^\zi(x)=0$ at points $x\in A_M$.

Taking local coordinates $(x^i, y^a)$ on $\sv(A)$ we may write $i_{\sv(A)}:
\sv(A)\hookrightarrow E$ as $i_{\sv(A)}(x^i, y^a)=(x^i, 0, 0, y^a, 0)$ and
$i_{\sv(A)}^*(x^i,0,\zx_0,\zx_a,\zx_\za)=(x^i,\zx_a)$, so
$$\sT i_{\sv(A)}^*(x^i,0,\zx_A,\dot{x}^j,0,y^B)=(x^i,\zx_a,\dot{x}^j,\dot{\zx}_a)\,.$$
For the adapted local coordinates
$(x^i,x^\zi,y^0,y^a,y^\za,p_i,p_\zi,\zx_0,\zx_a,\zx_\za)$ in $\sT^*E$, the map $\ze$
reduced to $(\sT^*E)_{\mid A}$ takes values in $\sT(E_{\mid A_M})$ (integrability) and
reads
$$\ze(x^i,0,1,y^a,0,p_I,\zx_A)=(x^i,0,\zx_A,\zr_e^j(x^i,0)y^e+
\zr_0^j(x^i,0),0,(c^D_{eB}(x^i,0)y^e+c^D_{0B}(x^i,0))\zx_D+\zs^I_B(x^i,0)p_I)\,.$$ Therefore
\bea\label{ac}&\sT
i_{\sv(A)}^*\circ\ze\left(x^i,0,1,y^a,0,p_I,\zx_A\right)=\\&\left(x^i,\zx_a,\zr_e^j(x^i,0)y^e+
\zr_0^j(x^i,0),\left(c^D_{eb}(x^i,0)y^e+c^D_{0b}(x^i,0)\right)\zx_D+\zs^I_b(x^i,0)p_I\right)\nn\eea and
\begin{eqnarray*}
&&\zL_L^A(x^i, y^a) =Ti_{\sv(A)}^*\circ\epsilon\left(x^i, 0,1,y^a, 0, \frac{\partial L}{\partial x^I}(x^j,0,1,
y^a, 0), \frac{\partial L}{\partial
y^A}(x^j,0, 1,y^a, 0)\right)\\
&&=\left(x^i, \frac{\partial L}{\partial y^b}(x^j, 0,1,y^a, 0), \rho^j_e(x^i,0) y^e+\zr^0(x^i,0),
\left({c}^D_{eb}(x^i,0) y^e+ c^D_{0b}(x^i,0)\right)\frac{\partial L}{\partial y^D}(x^j, 0,1,y^a,
0)\right.\\&&\left.+\sigma^I_b(x)\frac{\partial L}{\partial x^I}(x^j,0,1, y^a, 0)\right)\,.
\end{eqnarray*}
Therefore, locally,  the nonholonomic Euler-Lagrange equations read:
\bea\label{nh1a}
& x^\zi=0,\quad y^0=1,\quad y^\za=0,\quad \frac{d x^j}{dt}=\rho^j_e(x^i,0)
y^e+\zr^j_0(x^i,0)\\
& \frac{\xd}{\xd t}\frac{\partial L}{\partial y^b}(x^i, 0,1,y^a, 0)=\label{nh1b}\\& \left({c}^D_{eb}(x^i,0)
y^e+ c^D_{0b}(x^i,0)\right)\frac{\partial L}{\partial y^D}(x^i, 0,1,y^a, 0)+\sigma^I_b(x^i,0)\frac{\partial
L}{\partial x^I}(x^i, 0,1,y^a, 0) \,.\nn\eea On the other hand, (\ref{nh1b}) means that $\zd L_b=0$ for all
$b$, i.e. $\zd L\in \sv(A)^0$.
\end{proof}
In the case of a Lie algebroid and linear constraints A=\sv(A) covering the whole $M$, when we have in
particular $A_M=M$, $\sigma^i_e= \rho^i_e$, the previous equations are precisely the nonholonomic equations
obtained in \cite{CLMM} (see Equations 3.8).

\medskip\noindent Again, there is an obvious version of the above picture for a time-dependent Lagrangian based
on (\ref{diag1}). In the nonholonomic case, however, we cannot restrict the Lagrangian
function to the constraint, except for the case which is, in fact, {\it holonomic}.

\medskip\noindent
{\bf 5. Holonomic constraints.} In the nonholonomic case we can restrict the Lagrangian $L$ to the constraint
$S$ if the geometric constraint is holonomic. Note however, that this does not imply automatically that the
corresponding vakonomic and nonholonomic Euler-Lagrange equations are the same, since the equations are not
precisely variational (they describe only sufficient conditions that the variational principle holds true) and
they are obtained in different ways. On the other hand, in the linear case holonomicity means that the vector
subbundle $S$ is closed with respect to the algebroid bracket. Since the constraints are assumed to be
integrable, for the canonical Lie algebroid $E=\sT M$ this means, in turn, that $S=\sT S_M$, so the
constraints are holonomic in the classical sense. More generally, assume that $A$ is an affine constraint,
i.e. $A$ is an affine subbundle in $E$.
\begin{theo}\label{tac} An affine constraint $A$ in a quasi-Lie algebroid $E$ is holonomic if and only
if the algebroid bracket of sections of $A$ is a section of $\sv(A)$.
\end{theo}
\begin{proof} Let us choose a basis of sections $e_i$ and the corresponding linear coordinates
$(x^a,y^i)$ in $E$ such that $A$ is locally defined by equations $y^i=0,\quad i>r+1$, and $y^{r+1}=1$ and let
$\zg(t)=(x(t),y(t))$ be an admissible path in $A$. Then, $y^{i}(t)=0$ for $i>r+1$ and $y^{r+1}=1$. Moreover,
$\zz$ is a vertical variation of $\zg$, $\zz(x(t),y(t))=f^i(t)\pa_{y^i}$, if and only if $f^i=0$ for $i>r$. In
view of (\ref{tlift1a}), $\zd_\zz\zg$ is tangent to $A$ only if
$$\frac{\xd
f^k}{\xd t}(t) +c^k_{ij}(x(t))y^i(t)f^j(t)=0$$ for $k>r$. But for any $k>r$
$$\frac{\xd
f^k}{\xd t}(t) +\sum_{i,j}c^k_{ij}(x(t))y^i(t)f^j(t)=\sum_{j\le r}c^k_{(r+1)j}(x(t))f^j(t)+\sum_{i,j\le
r}c^k_{ij}(x(t))y^i(t)f^j(t)\,.$$ As $y^i(t),f^j(t)$ for $i,j\le r$ are arbitrary, $c^k_{ij}=0$ for $k>r$ and
$i\le r+1$, $j\le r$. Since $c^k_{ij}=-c^k_{ji}$, they vanish also for $k>r$ and for all $i,j\le r+1$. This
means that the bracket of local sections $[e_i,e_j]_\ze$, $i,j\le r+1$ belongs to the span of $\{
e_1,\dots,e_r\}$, i.e. is a section of $\sv(A)$. But sections of $A$ are of the form $e_{r+1}+\sum_{i\le
r}\zf_i(x)e_i$, so their brackets are sections of $\sv(A)$. The converse is obvious.
\end{proof}
\noindent According to the terminology of \cite{GGU1,GGU2,GGU3},
one can say that affine holonomic constraints in a Lie algebroid
are {\it Lie affgebroids}. A correct geometric description of
time-dependent systems and other systems, based on the idea of Lie
affgebroid was first proposed in \cite{SMM,MMS} and developed in
\cite{GGU1,GGU2,GGU3}.

\medskip\noindent
{\bf 6. Affine holonomic constraints -- geometric approach.}

If $A$ is a holonomic affine constraint, then, using local coordinates as above, we can
prove analogously to the proof of Theorem \ref{tac} that $c^D_{eb}(x^i,0)$ and
$c^D_{0b}(x^i,0)$ can be non-zero only for $D$ indexing a section of $\sv(A)$,
symbolically $D=d$, and that $\zs^\zi_b(x^i,0)=0$. Now, using the local form (\ref{ac})
of $\sT i_{\sv(A)}^*\circ\ze$, we conclude that $\sT i_{\sv(A)}^*\circ\ze$ vanishes on
the annihilator of $\sT A$. Hence, $\sT i_{\sv(A)}^*\circ\ze$ defines a map
$\ze^A:\sT^\ast A\ra\sT\sv(A)^\ast$ and the diagram (\ref{nhc}) reduces to the
following:
\be\label{nhc2} \xymatrix{ \sT^\ast A\ar[rrr]^{\ze^A}  &&&
\sT \sv(A)^\ast\ar[d]^{\zt_{\sv(A)^\ast}}\\
 A\ar[rrr]^{\zl_L^A}\ar[u]^{\xd L}\ar@{.>}[rrru]^{\zL_L^A}\ar[rrr]^{\zl_{L}}
 &&& \sv(A)^\ast}
\ee
This time, however, only the restriction of $L$ to $A$ does matter. The phase space is $\sv(A)^\ast$, the
phase dynamics is implicitly defined as $\zL^A_L(A)=\ze^A\circ\xd L(A)\subset\sT \sv(A)^\ast$, and the
Euler-lagrange equation for a curve $\zg$ in $A$ reads
$$\zL^A_L\circ\zg=\st(\zl_L^A\circ\zg)\,.$$
In local coordinates
\bea\label{nh1a2}
& x^\zi=0,\quad y^0=1,\quad y^\za=0,\quad \frac{d x^j}{dt}=\rho^j_e(x^i,0)
y^e+\zr^j_0(x^i,0)\\
& \frac{d}{dt}\left(\frac{\partial L}{\partial y^b}(x^i,y^a)\right)=\label{nh1b2}\\& \left({c}^d_{eb}(x^i,0)
y^e+ c^d_{0b}(x^i,0)\right)\frac{\partial L}{\partial y^d}(x^i, y^a)+\sigma^j_b(x^i,0)\frac{\partial
L}{\partial x^j}(x^i, y^a) \,.\nn
\eea
The above equations (canonically reduced to $A$) are exactly the Euler-Lagrange equations for a (Lie)
affgebroid obtained in \cite{GGU4, IMPS}. One can say that Geometrical Mechanics on a (Lie) affgebroid is just
Geometrical Mechanics on (Lie) algebroid with a holonomic affine constraint.


\bigskip
\noindent Katarzyna Grabowska\\
Division of Mathematical Methods in Physics \\
                University of Warsaw \\
                Ho\.za 69, 00-681 Warszawa, Poland \\
                 {\tt konieczn@fuw.edu.pl} \\\\
\noindent Janusz Grabowski\\Polish Academy of Sciences\\Institute of
Mathematics\\\'Sniadeckich 8, P.O. Box 21, 00-956 Warszawa,
Poland\\{\tt jagrab@impan.gov.pl}\\\\

\end{document}